\documentclass[reprint,aps,prd,nofootinbib,a4paper,10pt,twocolumn,superscriptaddress]{revtex4-2}
\usepackage{hyperref} 
\usepackage{changes}
\usepackage{graphicx,subfigure}
\usepackage[section]{placeins}
\usepackage[titletoc]{appendix}
\usepackage{xcolor}
\usepackage{dsfont}
\usepackage{bm}
\usepackage{mathtools} 
\usepackage{amsfonts}
\usepackage{enumerate}
\hypersetup{
     colorlinks   = true,
     citecolor    = cyan,
     linkcolor = pink
     }
\DeclareMathAlphabet{\mathpzc}{OT1}{pzc}{m}{it}

\newcommand{\sayy}[1]{`#1'}
\providecommand{\href}[2]{#2}

\def\be{\begin{equation}}
\def\ee{\end{equation}}
\def\bea{\begin{eqnarray}}
\def\eea{\end{eqnarray}}

\def\sig{\sigma}

\def\la{\langle}
\def\ra{\rangle}

\def\obs{\mathcal{O}}
\def\emi{\mathcal{E}}

\definecolor{MyB}{rgb}{0.1,0.1,1.0}

\definecolor{mygreen}{rgb}{0,0.5,0}

\definecolor{haypink}{rgb}{0.7,0,0.7}

\begin{document} 

\title{Redshift drift in a universe with structure III: Numerical relativity}  

\author{Sofie Marie Koksbang} 
\email{koksbang@cp3.sdu.dk}
\affiliation{CP$^3$-Origins, University of Southern Denmark, Campusvej 55, DK-5230 Odense M, Denmark}

\author{Asta~Heinesen}
\email{asta.heinesen@nbi.ku.dk}
\affiliation{Niels Bohr Institute, Blegdamsvej 17, 2100 Copenhagen, Denmark}  
\affiliation{Univ Lyon, Ens de Lyon, Univ Lyon1, CNRS, Centre de Recherche Astrophysique de Lyon UMR5574, F--69007, Lyon, France} 
\author{Hayley J. Macpherson}
\email{hjmacpherson@uchicago.edu}
\affiliation{Kavli Institute for Cosmological Physics, The University of Chicago, \\5640 South Ellis Avenue, Chicago, Illinois 60637, USA\\}
\affiliation{NASA Einstein Fellow}

\begin{abstract} 
Measurements of the cosmic redshift drift -- 
the change in redshift of a source over time --- will enable independent detection of cosmological expansion thanks to the immense precision soon reached by new facilities such as the Square Kilometer Array (SKA) Observatory and the Extremely Large Telescope (ELT). 
We conduct the first ever redshift drift computations in fully relativistic cosmological simulations, with the simulations performed with the Einstein Toolkit.
We compute the redshift drift over the full skies of 50 synthetic observers in the simulation. We compare all-sky averages for each observer---and across all observers---to the Einstein-de Sitter (EdS) model which represents the large-scale spatially-averaged space-time of the simulation.
We find that at $z\approx0.2$ the mean redshift drift across the sky for all 50 observers deviates from the EdS prediction at the percent level, reducing to $\sim0.1\%$ by $z\approx 1$. However, fluctuations in the redshift drift across the sky are $\sim$10--30\% at $z\approx 0.1$ and a few percent at $z\approx 0.5$. Such fluctuations are large enough to potentially exceed the expected precision of upcoming redshift drift measurements. 
Additionally, we find that along 0.48\% of the light rays, the redshift drift becomes temporarily positive at very low redshift of $z\lesssim 0.02$. 
This  occurs despite our simulation data being based on a matter-dominated model universe. 
By including a cosmological constant, we do expect a slower growth of structures than in the leading-order EdS space-time, and this may reduce the anisotropy over the observers's skies, although we generally expect our results to hold as order-of-magnitude estimates. 
Redshift drift is arguably one of the most important measurements to be made by next-generation telescopes. Our results collectively serve as preparation for interpreting such a measurement in the presence of realistic cosmic structures.

\end{abstract}
\keywords{Redshift drift, relativistic cosmology, observational cosmology} 

\maketitle

\section{Introduction}
One of the most exciting opportunities for cosmologists this century is the direct measurement of the expansion of the Universe. This can, for instance, be achieved through the \emph{redshift drift} which is the very slight change in the redshift of a source as observed
over time \citep{Sandage,Mcvittie}. Due to its expected small amplitude, it has only recently become feasible to build telescopes that will eventually be able to detect this signal. Such a measurement should become possible for the first time with telescopes such as the Square Kilometer Array (SKA) Observatory, which is anticipated to measure the redshift drift out to redshifts of $z\approx 1-2$ \citep{SKA, SKA2}.

The standard cosmological model rests on the use of the Friedmann-Lema\^itre-Robertson-Walker (FLRW) models---from theoretical predictions to observational analyses. 
In FLRW models, the redshift drift, $\delta z$, of a comoving source measured by a comoving observer is 
\begin{align}
\begin{split}
    \delta z &= \delta t_0 \left[(1+z)H_0 - H(z)\right], 
\end{split} 
\label{eq:flrwdrift}
\end{align}
where $\delta t_0$ denotes the proper-time interval of observation and $H(z)$ is the Hubble parameter. If the Hubble constant $H_0$ is known from other cosmological data, an observation of the redshift drift gives a direct measure of the Hubble parameter as a function of redshift. 
Moreover, because redshift drift reflects the change in expansion over time, measuring the redshift drift amounts to directly measuring whether the Universe is currently undergoing decelerated or accelerated expansion. 
Other cosmological probes typically rely on indirect measurements of the cosmic expansion by, e.g., determining the slope of the distance-redshift relation from supernovae. Redshift drift is the first probe that will allow us to \textit{directly measure} the acceleration of the Universe. 

Forthcoming measurements of the redshift drift represent a significant landmark for cosmology. We therefore must ensure that our analytical tools are sufficient to correctly interpret future redshift drift observations. Cosmological observations are known to be impacted by the presence of structures in the Universe, i.e., the deviation from the pure FLRW description of space-time.
The effect of the large-scale structures (LSS) on the redshift drift has been studied using a variety of exact inhomogeneous cosmological models \citep{2011PhRvD..83d3527Y, LTB2, stephani, Szekeres1, Szekeres2, Koksbang_2022, Koksbang:2015ctu, Wiltshire_2009} as well as anisotropic models \citep[e.g.][]{Bianchi1, Bianchi2}. The latter has become increasingly relevant with the observational support of large-scale anisotropies \citep[e.g.][]{anisotropy2, anisotropy3, anisotropy4, anisotropy5, anisotropy6, anisotropy7, anisotropy8}, though see, e.g., \cite{Ferreira:2021wr,Darling:2022ue,Horstmann:2022,Akarsu:2023ud} which find consistency with standard (isotropic) cosmology. 
See \citet{anisotropy1} for a recent review of such tensions.
Fluctuations in the redshift drift have also been studied analytically in terms of perturbation theory
\citep{Bianchi2, pert_durrer, pert1, pert2}, in Newtonian $N$-body simulations \citep{Koksbang_2023}, and in exact geometrical formulas for general-metric theories with arbitrary structures \citep{Korzynski:2017nas,Heinesen:2020pms, Heinesen:2021nrc, Heinesen:2021qnl}. 

There are at least three questions worth pursuing {in the theoretical and numerical study of redshift drift measurements, namely understanding (i) how local structures affect the redshift drift signal from individual sources, (ii) how structures affect the mean redshift drift signal upon averaging over many lines of sight, and (iii) whether the redshift drift signal can be used to develop observational tests of, e.g., cosmic inhomogeneity and/or anisotropy. 
In this work, we will focus on (i) and (ii) which are important for understanding potential biases 
in future redshift drift data from, e.g., SKA. 

The studies \cite{Koksbang:2019glb, Koksbang:2020zej}
showed that inhomogeneities can have a significant effect on the redshift drift when 
cosmic backreaction is non-negligible  \citep[see][for reviews of cosmic backreaction]{bcreview1, bcreview2, bcreview3}. 
On the other hand, studies based on models with negligible backreaction indicate that local structures along light paths only lead to small fluctuations in the redshift drift \cite{Koksbang_2023,Koksbang_2022, Bianchi2, pert_durrer, pert1, pert2, Koksbang:2015ctu}. 
Large inhomogeneities---such as a giant void---have been shown to have a significant effect on the measured redshift drift \cite{Koksbang:2015ctu, 2011PhRvD..83d3527Y, LTB2, Mishra:2012vi, Mishra:2014vga}. 

In this work, we will use numerical relativity to investigate redshift drift. Our simulations of large-scale structure formation are performed with full general relativistic evolution, under the assumption of a dust universe, beginning with small initial fluctuations inspired by the cosmic microwave background (CMB).
We perform ray-tracing within the same simulations without making any simplifying assumptions of an FLRW background or a perturbative expansion. This allows us to carry out realistic estimates of inhomogeneities on the redshift drift measurements for synthetic observers in the simulations.  
In Section~\ref{sec:model_setup} we describe our model setup including the simulation data and method for computing the redshift drift. We present our results in Section~\ref{sec:results} and give a summary and final comments in Section~\ref{sec:conclusion}.

\vspace{5pt} 
\noindent
\underbar{Notation and conventions:}
We use the Einstein summation convention and use the signature convention $(- + + +)$ for the spacetime metric $g_{\mu \nu}$, where Greek letters denote space-time indices. The covariant derivative, $\nabla_\mu$, is the Levi-Civita connection. A subscripted comma followed by an index indicates partial derivatives with respect to the indicated coordinate. 
Round brackets around indices denote symmetrization while square brackets denote anti-symmetrization.
We use $c = 1 = G $ throughout. 


\section{Model setup and light propagation} \label{sec:model_setup} 

\subsection{Numerical relativity simulations} 

The simulation data we use in this work is obtained using the open-source numerical relativity (NR) code the Einstein Toolkit\footnote{\url{https://einsteintoolkit.org}} \citep[ET;][]{Loffler:2012,Zilhao:2013}, which has been adapted and used for cosmological simulations of large-scale structure formation in recent years \citep[e.g.][]{Bentivegna:2016,Bentivegna:2017a,Macpherson:2017,Macpherson:2019a}. Initial conditions for the simulations are generated using \texttt{FLRWSolver} \citep{Macpherson:2017}, which provide linear perturbations about an Einstein–de Sitter (EdS) background cosmology. The initial fluctuations are drawn as random realisations of the matter power spectrum at the surface of last scattering as output from \texttt{CLASS}\footnote{\url{http://class-code.net}} \citep{Lesgourgues:2011} using $h=0.45$---where the Hubble constant is $H_0=100\, h$ km/s/Mpc---and otherwise default Planck parameters. 
We remove all power in our initial conditions below a scale corresponding to $\sim 10$ grid cells in each simulation in order to minimise the numerical error associated with under-sampling small-scale modes. 

We evolve the initial conditions using the 3+1 Baumgarte-Shapiro-Shibata-Nakamura (BSSN) formalism of NR from $\bar{z}=1000$ to $\bar{z}\approx 0$, where $\bar{z}$ is the redshift associated with the initial background EdS model. Note that this redshift is distinct from the redshifts computed along individual light rays with our ray tracing code (as described in Section~\ref{subsec:mescaline}), which are not restricted to follow the EdS model. 
The cosmological fluid is modelled as a perfect dust fluid with negligible pressure with respect to matter density, i.e. $P\ll\rho$, which has been shown to be sufficient to match an analytic EdS space-time evolution \citep{Macpherson:2017}. 
Further computational details---such as gauge choices and specific ET thorns used in the evolution---are as discussed in \citet{Macpherson:2019a}. 
In this work, we use two simulations with cubic box lengths of $L=3072\,h^{-1}$ Mpc and $1536\,h^{-1}$ Mpc with numerical grid resolutions of $N=256$ and $N=128$, respectively (where the total number of grid cells is $N^3$). Both simulations adopt periodic boundary conditions. These simulations sample the same minimum physical scale but have different numerical resolutions, and they are thus suitable for rough tests of
the numerical convergence of our results.
We present such a test in Appendix~\ref{app:128}, while we will present the results from the highest resolution simulation ($N=256$) in the main text. 

The simulation setup we use has been shown to exhibit small cosmic backreaction in the context of global spatial averages on the simulation hypersurfaces. Specifically, the averaged cosmological energy densities match the initial EdS background model predictions to within $\sim 2\%$ \citep{Macpherson:2019a}. 
We thus will compare the results we obtain for variance in the redshift drift to previous calculations based on analytic and exact solutions which exhibit negligible backreaction (\cite{Koksbang_2023} and the appendix of \cite{Koksbang:2015ctu}).

\subsection{Light propagation}\label{subsec:mescaline}

We calculate observables in our simulations using the post-processing code \texttt{mescaline} \cite{Macpherson:2023}. 
We place a set of present-epoch observers at random positions in the simulation domain and choose lines of sight which isotropically sample the sky for each observer. For each line of sight we generate the corresponding null geodesic and trace it backwards in time.

\texttt{Mescaline} solves the geodesic and geodesic deviation equations using the simulated metric tensor and the chosen initial data (observer position and directions of observation). The observers are chosen to be co-moving with the fluid flow in the simulation, i.e., their 4--velocity $u^\mu$ coincides with the 4--velocity of the fluid at their location in the simulation domain. All ``sources'' along each line of sight are also co-moving with the fluid flow in the simulation, such that the photon energy (and subsequently the redshift) along each geodesic is defined as $E\equiv - u^\mu k_\mu$ where $u^\mu$ is the 4--velocity of the fluid at the position of the geodesic, and $k^\mu$ is the tangent vector of the null geodesic.
For details of the ray-tracing procedure, including tests of the code and specifics on generating initial data, see \citet{Macpherson:2023}. 

For this work, we choose 50 randomly placed observers with lines of sight coinciding with the directions of \texttt{HEALPix}\footnote{\url{http://healpix.sourceforge.net}} pixels with $N_{\rm side}=8$ \citep{Gorski:2005}. 
We propagate light rays from each observer position until the redshift reaches $z\approx 1$.

\subsection{Redshift drift computation} \label{subsec:light}

\texttt{Mescaline} computes the null geodesic tangent vector, redshift, angular diameter distance, and Jacobi matrix along each geodesic \citep[details of these calculations can be found in][]{Macpherson:2023}.
In this section we detail the redshift drift computation we have introduced into \texttt{mescaline}.
We have used the multipole decomposition of the redshift drift introduced in \cite{Heinesen:2020pms,Heinesen:2021nrc,Heinesen:2021qnl}. We thus consider an observer moving with 4--velocity $u^\mu$ with corresponding proper time $\tau$ satisfying $\dot{\tau} = 1$, with the overdot representing parallel transport along $u^\mu$, i.e., $\dot{} \equiv u^\mu \nabla_\mu$.
For the corresponding congruence we have the following decomposition: 
\bea 
    \label{def:expu}
    && \nabla_{\nu}u_\mu  = \frac{1}{3}\theta h_{\mu \nu }+\sig_{\mu \nu} + \omega_{\mu \nu} - u_\nu a_\mu  \ , \nonumber \\ 
    && \theta = \nabla_{\mu}u^{\mu} \, ,  \quad \sig_{\mu \nu} = h_{ \la \nu  }^{\, \beta}  h_{  \mu \ra }^{\, \alpha } \nabla_{ \beta }u_{\alpha  }  \, , \nonumber \\ 
    && \omega_{\mu \nu} = h_{  \nu  }^{\, \beta}  h_{  \mu }^{\, \alpha }\nabla_{  [ \beta}u_{\alpha ] }   \, , \quad  a^\mu = \dot{u}^\mu \,  , 
\eea 
where $h_{ \mu \nu } = u_{ \mu } u_{\nu } + g_{ \mu  \nu }$ projects onto the spatial hypersurfaces orthogonal to the 4--velocity of the observer; on which it also plays the role of an adapted spatial metric tensor. In the above, $\omega_{\mu\nu}$ is the vorticity, $\sigma_{\mu\nu}$ is the shear, $\theta$ is the expansion, and $a_{\mu}$ is the 4--acceleration of the congruence.

We introduce $e^\mu = u^\mu - \frac{1}{E} k^\mu$ which is the viewing angle/spatial direction vector of observation of the light ray as seen by an observer in the frame of $u^\mu$. 
Thus, the photon 4--momentum can be decomposed as
\begin{equation}\label{eq:kmu_decomp}
    k^\mu = E (u^\mu - e^\mu).
\end{equation}
Since $e^\mu$ 
represents the position of a given source on the observer's sky, we introduce the \emph{position drift} of the source as
\bea
    \label{positiondrift}
    \kappa^\mu =  h^{ \mu }_{\; \nu }   \dot{e}^\nu \, , 
\eea  
where the source is kept fixed under the time derivative. 
We then write the redshift drift as an integral along the light ray according to 
\bea \label{eq:zdrift_integral}
    \label{redshiftdriftint}
    \frac{d z}{d \tau} \Bigr\rvert_{\obs} = E_\emi \! \! \int_{\lambda_\emi}^{\lambda_\obs} \! \! d \lambda \,   \Pi     \, , \qquad z = \frac{E_\emi}{E_\obs} - 1,
\eea  
where $\lambda$ is an affine parameter along the light ray and $\obs$ and $\emi$ represent the points of observation and emission, respectively. Using the results of \citet{Heinesen:2021qnl} we can write the integrand in full generality as
\bea
    \label{Pimultikappa}
    \hspace*{-0.65cm} \Pi &=&  -  \kappa^\mu \kappa_\mu  + \Sigma^{\it{o}}    +  e^\mu \Sigma^{\bm{e}}_\mu    +       e^\mu   e^\nu \Sigma^{\bm{ee}}_{\mu \nu} + e^\mu   \kappa^\nu \Sigma^{\bm{e\kappa}}_{\mu \nu},
\eea  
where the coefficients are given by
\bea
    \label{Picoefkappa}
    && \Sigma^{\it{o}} =  - \frac{1}{3} u^\mu u^\nu R_{\mu \nu}     + \frac{1}{3}D_{\mu} a^{\mu} + \frac{1}{3} a^\mu a_\mu    \, , \nonumber  \\   
    &&  \Sigma^{\bm{e}}_\mu  =   - \frac{1}{3}  \theta a_\mu    -   a^{ \nu} \sigma_{\mu \nu}  + 3 a^{ \nu} \omega_{\mu \nu}    - h^{\nu}_{\mu} \dot{a}_\nu \, ,  \nonumber  \\   
    && \Sigma^{\bm{ee}}_{\mu \nu} =     a_{\la \mu}a_{\nu \ra }  + D_{  \la \mu} a_{\nu \ra }    -  u^\rho u^\sigma  C_{\rho \mu \sigma \nu}   -  \frac{1}{2} h^{\alpha}_{\, \la \mu} h^{\beta}_{\, \nu \ra}  R_{ \alpha \beta }  \,   , \nonumber  \\   
    && \Sigma^{\bm{e\kappa}}_{\mu \nu} =  2 (\sigma_{\mu \nu}  - \omega_{\mu \nu}   )  \, . 
\eea  
Here, $R_{\mu \nu}$ is the Ricci tensor, $C_{\rho \mu \sigma \nu}$ is the Weyl tensor, and $D_{\mu} a_{\nu} ,   =  h_{ \nu }^{\, \alpha }  h_{ \mu }^{\, \sigma } \nabla_\sigma  a_{\alpha}$ is the spatial covariant derivative of the fluid 4--acceleration.
 
We introduce some simplifications to the above equations both to make them easier to implement into \texttt{mescaline} and to ensure reasonable computation time. 
Since the simulation models the cosmic fluid as dust,  
the 4--acceleration and vorticity in the simulation are subdominant in \eqref{Picoefkappa} above.
 
We hence ignore all terms including the vorticity and 4--acceleration. 
Likewise, we neglect the Ricci term in $\Sigma^{\bm{ee}}_{\mu \nu}$, as it represents the anisotropic stress of the fluid which is conditioned to be zero in the perfect-fluid assumption of the simulations. 

Finally, we also neglect terms including the
position drift $\kappa^\mu$ defined in \eqref{positiondrift}. 

This approximation is more subtle, as there are no constraints built into the ET simulations that \emph{a priori} forces these terms to be small.
An earlier study \cite{Koksbang_2023} of Lema\^itre-Tolman-Bondi \citep[LTB;][]{lemaitre, tolman, bondi} models indicates that by dropping terms containing $\kappa^\mu$, one neglects contributions of the order $\lesssim 1 \%$ of the total redshift drift signal for light propagation out to $z\sim 0.02$, corresponding to the size of the void in the particular model. We expect this to represent an upper bound on the position drift for similar redshifts (with a relative decrease towards higher redshifts) in models with more modest structures and a notion of statistical homogeneity. 
In linearised FLRW models, typical values of position drift at cosmological distances are of order $v H$, where $v$ is the size of the velocity of the source relative to the Poisson frame and $H$ is the Hubble parameter \cite{Rasanen:2013swa}. Since $\kappa^\mu$ enters the redshift drift formula either in terms involving its quadrature or in terms where it is multiplying shear/vorticity, the contributions involving $\kappa^\mu$ are expected to be second order in regions where density contrasts may be considered perturbative. 
These considerations, together with the upper bound estimates provided by  
\cite{Koksbang_2023}, motivate us to neglect terms involving $\kappa^\mu$ in our present investigations.

Under the above approximations, \eqref{Pimultikappa} 
reduces to
\bea\label{eq:simplified_Pi}
    \label{Pisimple}
    \hspace*{-0.65cm} \Pi 
    &\approx & - \frac{1}{3} u^\mu u^\nu R_{\mu \nu} -        e^\mu   e^\nu     u^\rho u^\sigma  C_{\rho \mu \sigma \nu}.
\eea 

We will in the following refer to these two terms as the Ricci and Weyl contributions, respectively.

We substitute \eqref{eq:simplified_Pi} into \eqref{eq:zdrift_integral} to obtain the following discretised integral-formula for the redshift drift
\begin{align}\label{eq:sum}
    \delta z = \delta \tau_\obs\,  E_\emi &\sum_{\lambda=\lambda_\obs}^{\lambda_\emi} \bigg( - \frac{1}{3} u^\mu u^\nu R_{\mu \nu} \\
    &-        e^\mu   e^\nu     u^\rho u^\sigma  C_{\rho \mu \sigma \nu} \bigg) \Delta \lambda,\nonumber
\end{align}
which is the computation 
we add into the \texttt{mescaline} ray tracer \citep{Macpherson:2023}. 
For all cases in this work we choose the observation period $\delta\tau_\obs=30$ years 
since it represents a typical value of the observation time interval expected for surveys \citep[see, e.g.][]{Liske:2008ph}, but note that $\delta \tau_\obs$ simply represents a scaling of the redshift drift signal. In relation to this, we note that the individual terms in eq. \ref{eq:sum} are typically of absolute order $10^{-4}- 10^{-2}$ up to the scaling by $\delta \tau_\obs$, energy and $d\lambda$. Thus, although the final redshift drift is roughly of order $10^{-10}$ (because we multiply with $\delta t_0\sim 10^{-8}$ at the very end), much lower absolute precision is required for the actual computations. We verify that the numerical noise for the individual terms along the rays is much smaller than the size of the terms by e.g. controlling that the null condition is fulfilled to high precision along the light rays and by verifying (in appendix \ref{app:LTBtest}) that our code can accurately reproduce known results for specific inhomogeneous metrics.

We calculate the affine interval $\Delta\lambda$ via the Jacobian 
\begin{align}
    \frac{dt}{d\lambda} &= -\frac{1}{\alpha}k^\mu n_\mu = \frac{E_n}{\alpha} , 
\end{align}

where $\alpha$ is the lapse function, $n_\nu=(-\alpha, {\bf 0})$ is the normal vector of the spatial hypersurfaces, $t$ is the simulation coordinate time, and $E_n$ is the photon energy in the simulation hypersurface frame. 

The Ricci and Weyl tensor components are calculated using existing routines in \texttt{mescaline} \citep{Macpherson:2019a,Macpherson:2023} and interpolated to the exact position of the geodesic. 
We set observers and emitters comoving with the fluid flow, i.e., we set $u^\mu$ to coincide with the 4--velocity of the fluid at the position of the geodesic in the simulation. We use the photon 4--momentum $k^\mu$---which is evolved using the geodesic equation in \texttt{mescaline}---and \eqref{eq:kmu_decomp} to calculate the direction vector $e^\mu$ along the geodesic.

\section{Results}\label{sec:results}


In this section, we will present results from propagating 
light rays through the simulation data with box length $L=3072\,h^{-1}$ Mpc and {a grid resolution of} $N = 256$. We consider 50 observers each with 768 lines of sight propagated until $z\approx 1$. 
We are interested in studying the variance in the redshift drift along different lines of sight for a particular observer, as well as the mean of the redshift drift across each observer's sky. Although not observable, we will also assess the mean across all observers at different positions in the simulation. In particular, we will assess whether these mean values coincide with the prediction from the initial background EdS model of the simulation. 

\begin{figure}
\centering
\includegraphics[scale = 0.5]{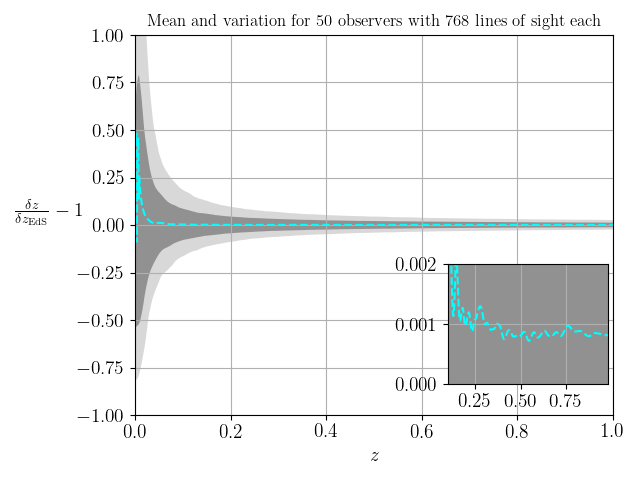}
\caption{Mean and fluctuations of the redshift drift along 768 random light rays each for 50 randomly placed present-time observers. The redshift drift is shown relative to the EdS redshift drift. The dark and light shaded regions show the 68.1\% and 95.4\% percentiles across all light rays, respectively. 
}
\label{fig:dz}
\end{figure}
In Figure~\ref{fig:dz} we show the redshift drift normalised to the EdS prediction. We show the mean and variance in terms of 68.1\% and 95.4\% percentiles 
when considering all lines of sight for all observers (38400 total lines of sight). These percentiles correspond to the $1-\sigma$ and $2-\sigma$ confidence intervals in the limit of a Gaussian distribution. 
For $z\gtrsim 0.03$, the mean of the redshift drift deviates from the EdS model by less than $0.2\%$. 
This is in good agreement with work using LTB models \citep{Koksbang:2015ctu} and Newtonian $N$-body simulations \citep{Koksbang_2023} in the sense that these studies similarly indicate that structures of realistic size have only small impact on the mean redshift drift.
The maximum fluctuations across all light rays we find are  $\sim 10\%$ at $z\approx 1$. The 95.4\% percentiles are $\sim\pm$2--3\% about the mean at this same redshift. The $\sim 10\%$ maximum fluctuations are more than an order of magnitude larger than the 
$<1\%$ variance found using Newtonian $N$-body simulations combined with a perturbative scheme \citep{Koksbang_2023}. 
This indicates that such an approximate procedure may be insufficient for quantifying fluctuations in the redshift drift. 
However, this difference in results could also reasonably be due to, e.g., the different physical resolutions as well as differences in modelling of the matter field: in our ET simulations there is a hydrodynamic model of the matter whereas the matter is simulated as a particle ensemble in \citet{Koksbang_2023} --- thus the latter necessitates a smoothing procedure. 

Notably, the maximum fluctuations obtained with the Newtonian simulation lie below the expected precision of SKA, however, the maximum fluctuations we find here are above the survey's expected precision of 1--10\% \cite{SKA}. Further, even our 68.1\% fluctuations in the redshift drift may lie above this level of precision and thus could be relevant to consider.

\begin{figure}
\centering
\includegraphics[scale = 0.5]{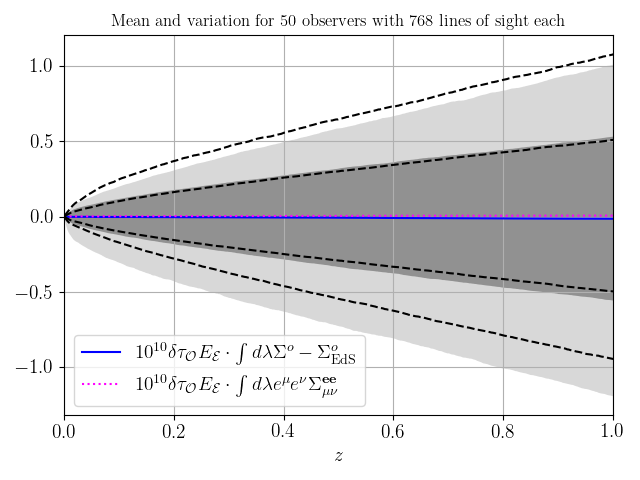}
\caption{Mean and fluctuations of the Weyl and Ricci contributions to the redshift drift for 50 
observers with 768 lines of sight each. The dark and light shaded area shows the 68.1\% and 95.4\% percentiles across all light rays 
for the Ricci component, respectively, while the same is shown for the Weyl component with dashed curves.
}
\label{fig:WeyldRicci}
\end{figure}
In Figure~\ref{fig:WeyldRicci} we show the two individual components of the redshift drift in \eqref{eq:simplified_Pi}. The Ricci component (the approximation to $\Sigma^{\it{o}}$) is shown relative to the EdS value (the EdS contribution is zero for the Weyl component). 
The mean value of the Ricci contribution is shown with a solid blue curve and the Weyl contribution is shown with a dashed pink curve.
The 68.1\% and 95.4\% confidence intervals across all light rays for all observers are shown as dark and light grey shaded regions for the Ricci and as dashed black curves for the Weyl contribution.
We find the mean value of these contributions is dependent on the numerical resolution of the simulation (see Appendix~\ref{app:128}). 
However, the order of magnitude of the variances we find is robust to resolution changes. Further, the slight skew of the Ricci contribution towards negative values and the Weyl contribution towards positive values is also consistent between resolutions. We also notice that the Weyl and Ricci contributions are not perfectly symmetric, indicating that they do not cancel each other perfectly along individual light rays. In comparison, these were found to cancel almost completely in the study based on $N$-body simulations combined with perturbation theory in \citet{Koksbang_2023}. From equation~15 in \citet{Koksbang_2023}, this indicates larger spatial derivatives of the metric components in our simulation setup with respect to those $N$-body simulations. 

\begin{figure}
\centering
\includegraphics[scale = 0.5]{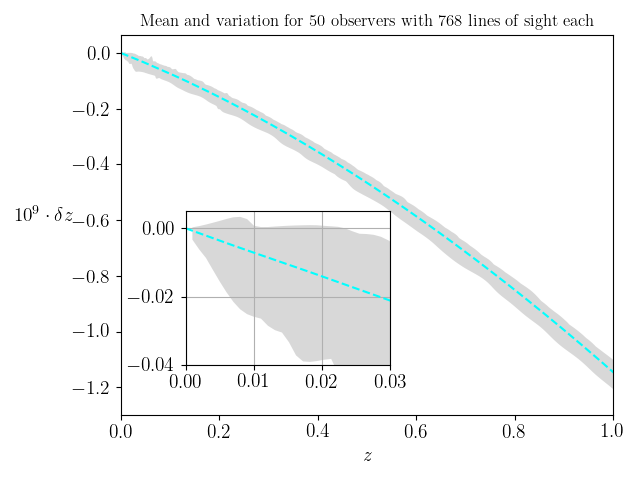}
\caption{The curve shows the mean of the redshift drift along 768 light rays for 50 observers. The grey shaded region shows the maximum to minimum fluctuations of $\delta z$ across all light rays. }
\label{fig:redshift_drift}
\end{figure}
Figure~\ref{fig:redshift_drift} shows the redshift drift averaged along 768 light rays for all 50 observers. The dashed curve shows the mean and the grey shaded region shows the maximum to minimum variance across all light rays. 
Most notably, the redshift drift along 185 light rays (corresponding to 0.48\% of the 38400 light rays) spread over 5 of the 50 observers is \emph{positive} at very low redshifts ($z\lesssim 0.01$). This is possible because the Weyl contribution to the redshift drift can be positive, and in some cases it dominates the total redshift drift signal at low redshifts. 
Physically, redshifts of $z\sim 0.01$ roughly correspond to a distance scale comparable to the largest walls/voids in the simulations. At low redshifts, the EdS background redshift drift is of order $z H_0$, while peculiar acceleration (in an estimate based on linearised perturbation theory) is of order $v H_0$ where $v$ is the size of the velocity of the source relative to the Poisson frame. Thus, roughly, we expect the peculiar acceleration contribution to be dominant when $v \gtrsim z$. Since velocities in the simulation are of order $\ll0.01$ in units of the speed of light, 
we expect some light rays to have positive redshift drift for $z\lesssim 0.01$ which is exactly what is seen in Figure~\ref{fig:redshift_drift}. The mean redshift drift is, however, always non-positive as expected since our simulations do not contain any dark energy component \citep{Heinesen:2021nrc}. We also note that positive redshift drift due to structures have earlier been found in LTB models in \cite{2011PhRvD..83d3527Y}, when placing an observer in a large central overdensity surrounded by a steep transition to a deep undersensity.

\begin{figure}
    \centering
    \includegraphics[scale = 0.5]{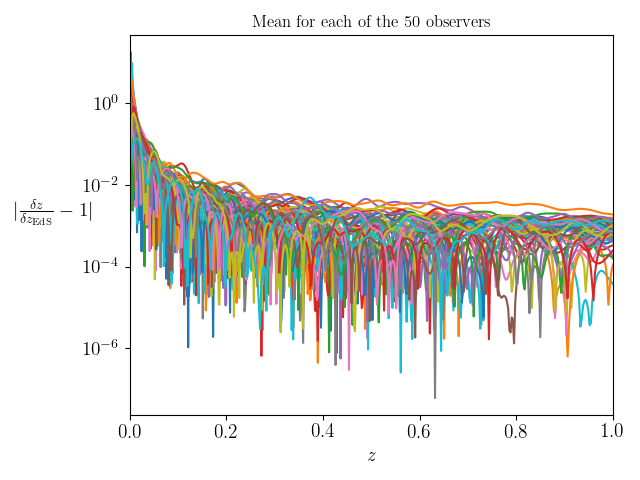}
    \caption{Mean redshift drift across 768 lines of sight for 50 observers as a function of redshift, $z$. We show the absolute value of $\delta z$ normalised by the background EdS prediction, $\delta z_{\rm EdS}$.}
    \label{fig:dz_all}
\end{figure}
Figure~\ref{fig:dz_all} shows the mean redshift drift averaged over 768 lines of sight for each of our 50 observers. The mean values all tend towards the EdS result at $z\rightarrow 1$. However, the deviation from the EdS prediction remains of order $\sim 1\%$ at $z\approx 0.2$. These deviations average out to $<0.1\%$ when we take the mean of all observers (see Figure~\ref{fig:dz}). Real observations are of course based on measurements from only a single observer, and the $\sim 1\%$ fluctuations could be of importance for measurements of high enough precision. We see a difference at the $\sim 1\%$ level at cosmological redshifts across observers in Figure~\ref{fig:dz_all}, and determining where our observations might lie in such a distribution is important for any future measurements approaching this level of accuracy. Measuring the redshift drift to percent-level precision may be feasible with Phase~2 of SKA \cite{SKA}.
Making use of, e.g., constrained cosmological simulations \citep[e.g.][]{Dolag:2023wf,Klypin:2003uh} combined with relativistic ray tracing could enable us to study these variances realistically in our local environment.

\subsection{Skymaps and power spectra}\label{subsec:skymaps}

In addition to the variances in the mean redshift drift for observers in different physical environments, the variance across individual lines of sight could also impact future measurements. 
In particular, many surveys do not cover the entire sky due to, e.g., the galactic centre or simply the survey footprint. Anisotropic variance of cosmological parameters is expected due to inhomogeneities, especially at low redshifts \citep[e.g.][]{Heinesen:2020bej,Macpherson:2021a}, and this naturally extends to the redshift drift \citep{Heinesen:2021qnl}. 

In our analysis so far, we used relatively low-resolution skies of $N_{\rm side}=8$ for the $N=256$ resolution simulation because of the increased computational cost of propagating many light rays for this data. With the lower-resolution simulation of $N=128$, we can use higher-resolution skies with $N_{\rm side}=32$ to enable us to study the angular distribution of the redshift drift in better detail. 
For this case, we use 10 randomly-placed observers and propagate light rays to $z\approx 0.5$. 
In Appendix~\ref{app:128} we show that the variance of $\delta z$ is robust to this change in numerical resolution. 

\begin{figure}
    \centering
    \includegraphics[width=\columnwidth]{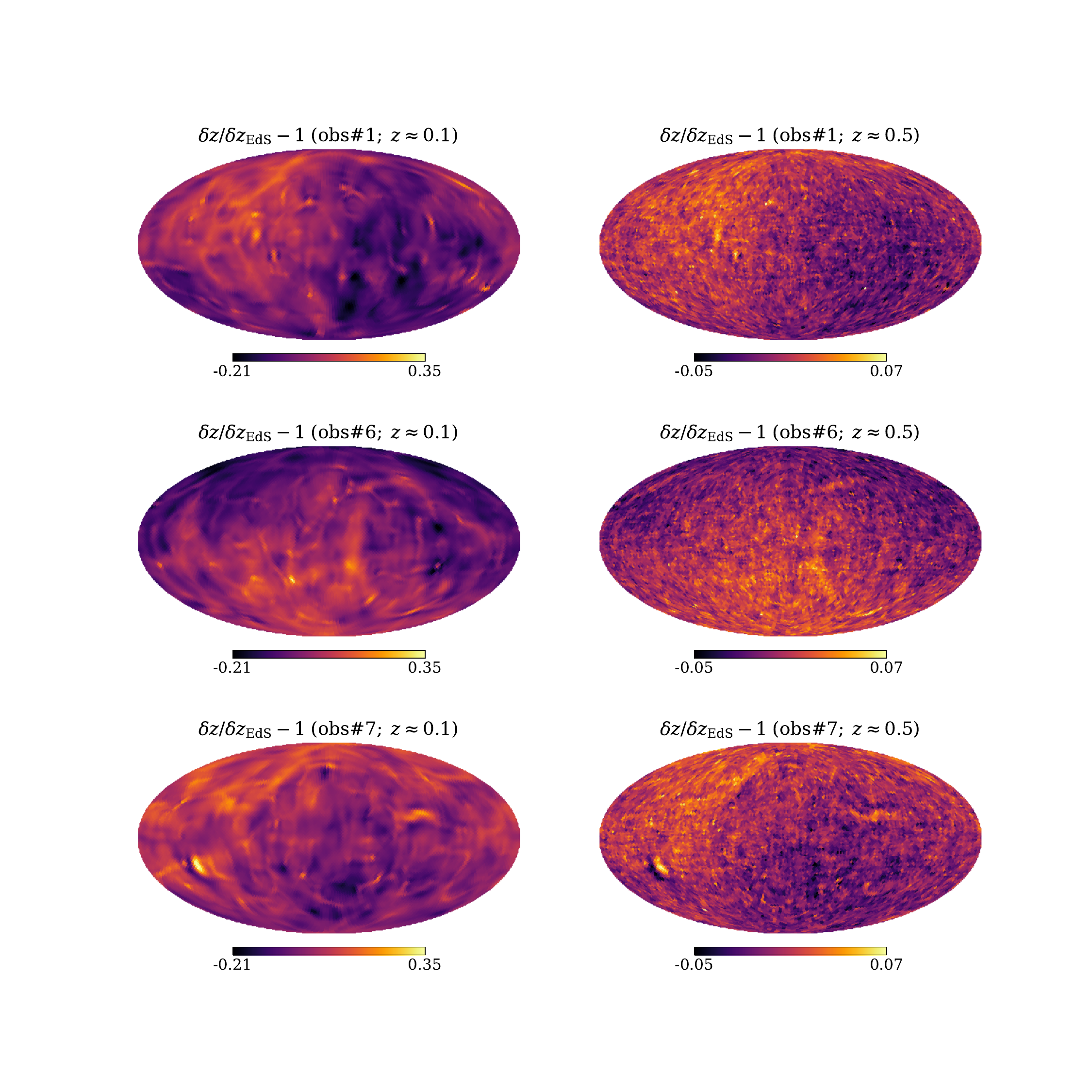}
    \caption{All-sky maps of the redshift drift, $\delta z$, relative to the background EdS value for three sample observers at two redshift slices $z\approx 0.1$ and $z\approx 0.5$. Each observer has $N_{\rm side}=32$ lines of sight in directions of \texttt{HEALPix} indices.}
    \label{fig:dz_skymaps}
\end{figure}
In Figure~\ref{fig:dz_skymaps} we show the all-sky redshift drift signal for three sample observers (panels; top to bottom) and for two redshift slices of $z\approx 0.1$ (left panels) and $z\approx 0.5$ (right panels). For each panel we are showing the redshift drift relative to the EdS value, $\delta z / \delta z_{\rm EdS} -1$. The EdS value is calculated using \eqref{eq:flrwdrift} with the mean redshift of each slice for each observer as $z$, the globally-averaged Hubble constant for the simulation as $H_0$, and the EdS relation for the Hubble parameter, $H(z)=H_0 (1+z)^{3/2}$}.  
At redshifts of $z\approx 0.1$, our observers see a variance in the redshift drift of tens of percent, with large-angle multipoles visibly dominating the signal. For $z\approx 0.5$, the sky-variance has reduced to several percent with respect to the EdS signal, with a visible dipole anisotropy present. 

\begin{figure}
    \centering
    \includegraphics[width=\columnwidth]{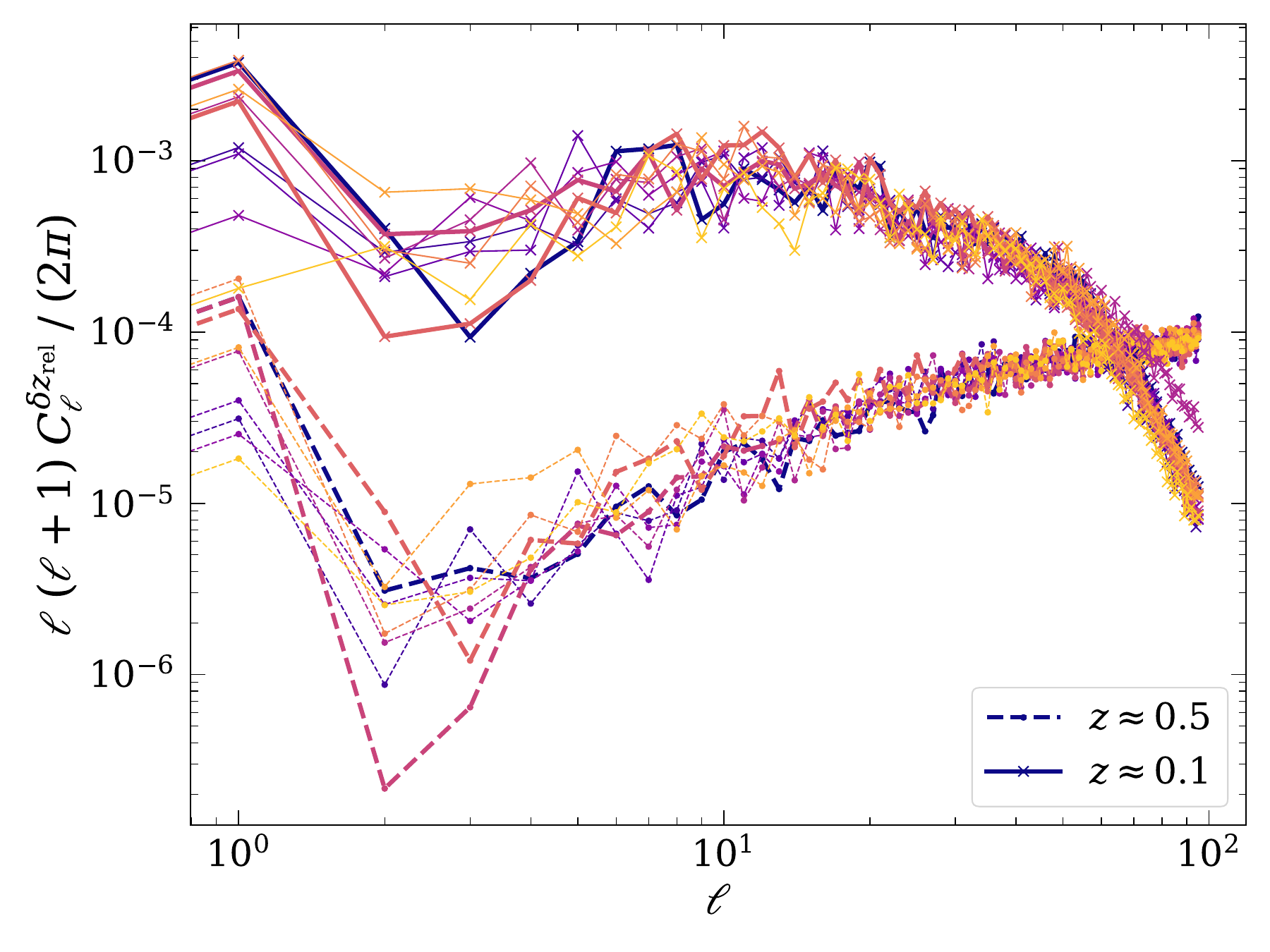}
    \caption{Angular power spectra of the redshift drift maps for all 10 observers at two redshift slices $z\approx 0.1$ (dashed curves) and $z\approx 0.5$ (solid curves). We show the power spectra of $\delta z_{\rm rel}\equiv \delta z / \delta z_{\rm EdS} - 1$. 
    Colors of each line are the same for individual observers between redshifts. Thick curves for both redshifts indicate the observers shown in the sky maps in Figure~\ref{fig:dz_skymaps}. 
    }
    \label{fig:dz_Cls}
\end{figure}
We can further explore the multipole contributions to the redshift drift signal by calculating the angular power spectra of the maps in Figure~\ref{fig:dz_skymaps} for all 10 observers. 
Figure~\ref{fig:dz_Cls} shows the angular power spectra, $C_\ell$, as a function of multipole $\ell$ for $\delta z_{\rm rel}\equiv \delta z / \delta z_{\rm EdS} - 1$. 
The maximum $\ell$ shown on the plot is $\ell_{\rm max}$: the smallest angular scale we can resolve for our ray-tracing maps with $N_{\rm side}=32$. Solid curves show the $C_\ell$ for all observers on a redshift slice of $z\approx 0.1$ and dashed curves of the same colour show the same observers for $z\approx 0.5$. Thick curves for both slices indicate the same observers shown in Figure~\ref{fig:dz_skymaps}. 
We can see the power in deviations from EdS reduces at almost all multipoles (except those which are close to $\ell_{\rm max}$) as we go to higher redshift. At $z\approx 0.5$, we see a relative increase in power in small angular scales with respect to lower redshift. However, the dipole remains a dominant contribution to the signal---an order of magnitude larger than the quadrupole for most observers. This is perhaps expected in our simulations because gradients of expansion (contribution to the dipole of $\delta z$) dominate over the local shear (contribution to the quadrupole) \citep{Heinesen:2021qnl,Macpherson:2021a}. 

We can quantify the amplitude of the low-$\ell$ multipoles of the angular power spectrum of the redshift drift, $C_{\ell}^{\delta z}$, relative to the monopole component. This amplitude is
\begin{equation}
    A_\ell=\sqrt{\frac{(2\ell+1)C_{\ell}^{\delta z}}{C^{\delta z}_{\ell=0}}}.
\end{equation} 
At $z\approx0.1$, we find that the mean dipole amplitude across the 10 observers in the $N=128$ simulation is $3.7\%$, and that the mean quadrupole amplitude is 1.1\%. At a higher redshift of $z\approx 0.5$, the mean dipole and quadrupole amplitudes are 0.8\% and 0.1\% of the monopole, respectively.
\newline\indent
We note that the power spectra for the relative redshift drift were  investigated in \cite{pert_durrer} within linearized perturbation theory. 
There are qualitative similarities between the power spectra shown here and those presented in \cite{pert_durrer}, such as the power spectra for $z \approx 0.1$ being above the $z\approx 0.5$ curve at low $\ell$, but vice versa at higher $\ell$. 
Quantitatively, there is a difference in scale between the power spectra in Figure~\ref{fig:dz_Cls} and those at similar redshifts in \cite{pert_durrer}, with the power spectra in \cite{pert_durrer} being significantly suppressed relative to ours. 
This is primarily because the expression for redshift drift used in \cite{pert_durrer} to calculate the power spectra is in fact not the full redshift drift. In particular, it neglects the term, $(1+z)H_0$, which is one of the two counteracting leading-order monopole terms in the FLRW expression. As a consequence their powerspectra of their \emph{normalised} redshift drifts $\delta z_{\rm rel}\equiv \delta z / \delta z_{\rm FLRW} - 1$ get suppressed, especially at small redshifts, where the suppression at $z\approx 0.1$ is more than two orders of magnitude relative to the power spectrum for the true redshift drift signal. After accounting for this issue with normalisation, the power spectra of \cite{pert_durrer} are largely consistent with our results. Any additional difference we expect to come from a difference in choice of background cosmology of the two studies, and towards high $l$ we also expect the non-linearities -- which are incorporated in our investigations but neglected in \cite{pert_durrer} -- to play a role. 
\newline\newline
To confirm that the power spectra presented in figure \ref{fig:dz_Cls} do not represent noise, we have compared the $z \approx 0.1$ power spectra with different values of $N$ and $N_{\rm side}$, i.e. different resolutions of the sky and simulation. The comparison (presented in appendix \ref{app:power_spectrum}) shows very similar behavior in all three cases.

\section{Caveats}\label{sec:caveats}

Here we outline some important caveats to the results we have presented in this work. These caveats should be carefully considered when physically interpreting our results. 

In this investigation, the cosmological constant is zero due to the fact that the ET does not currently include a cosmological constant in Einstein's equations. 
Therefore, inhomogeneities in our simulations are enhanced with respect to a model universe with a cosmological constant. As a result, the fluctuations in the redshift drift that we find are most likely larger than would be expected in 
simulations based on the $\Lambda$CDM model.

Further, we did not incorporate the position drift of the source into our calculation of the redshift drift in the simulations. 
A combination of earlier studies \cite{Koksbang_2022, Koksbang_2023} of the effect within extreme LTB structures combined with arguments from perturbation theory allows us to make plausible that this effect is small.
Computing the position drift is significantly more complicated than computing the Ricci and Weyl components which we have focused on here. Including the position drift would involve solving an additional 20 differential equations along the light paths, and given that the effect is likely small we neglect it in our analysis. 
Alternatively, future work might consider using the code BigOnLight \citep{Grasso:2021iwq} to study the redshift drift in ET simulations.

\section{Summary and conclusion} \label{sec:conclusion} 
We have studied redshift drift along light rays in simulation data sets generated using the Einstein Toolkit and \texttt{mescaline}. 
The simulation data represents fully general-relativistic 
matter-dominated cosmologies with initial fluctuations based on standard perturbation theory. 

We find that the mean redshift drift of individual observers is very close to the EdS model, typically of order $\lesssim 1\%$ in the redshift interval $0.2 \lesssim z \lesssim 1.0$, which is consistent with large-scale spatial averages in the simulations. The maximum fluctuations around the mean
are above 10\% at $z = 0.5$ and still over 5\% at $z = 1$. 
These fluctuations represent the maximum and minimum across many light rays and many observers. In reality, most light rays will not experience this level of variance. The 95.4\% confidence interval is $\pm 2.5\%$ fluctuations about the mean and the 68.5\% confidence interval is $\pm1$\% at $z\approx 1$. 

We compute the redshift drift as an integral over two separate contributions from the Weyl and Ricci tensors. The fluctuations about the mean of the Weyl and Ricci components both increase with redshift. However, on average, the two components largely cancel with each other, yielding a mean redshift drift signal which deviates from the EdS signal at the sub-percent level. 

We have shown that the redshift drift becomes positive at very low redshifts along 0.48\% of the light rays. This is made possible by a slight dominance of the Weyl contribution at low redshift. 
The positivity of the redshift drift signal is interesting because it is a signature for an accelerating space-time in FLRW geometry. However, this only occurs at low redshift where inhomogeneity is more significant. We find no \sayy{apparent dark energy signature} at cosmological redshifts in our simulations.

We also investigated the variance of the redshift drift across the sky for a set of 10 observers with well-sampled skies. Such a variance is of relevance for surveys with incomplete sky coverage. We found a $\sim$10--30\% sky-variance at $z\approx 0.1$ which reduces to several percent by $z\approx 0.5$. Studying the angular power spectra at these redshifts, we find the dipole to be a dominant contribution to the signal for most observers. 
The sky-variance we find here is of same order of magnitude as the anticipated precision for future surveys intending to measure the redshift drift at percent precision such as Phase~2 of SKA \cite{SKA}.

\vspace{6pt} 
\begin{acknowledgments}
SMK is funded by VILLUM FONDEN, grant VIL53032. AH acknowledges funding from the Carlsberg Foundation and the European Research Council (ERC) under the European Union’s Horizon 2020 research and innovation programme (grant agreement ERC advanced grant 740021–ARTHUS, PI: Thomas Buchert).
Support for HJM was provided by NASA through the NASA Hubble Fellowship grant HST-HF2-51514.001-A awarded by the Space Telescope Science Institute, which is operated by the Association of Universities for Research in Astronomy, Inc., for NASA, under contract NAS5-26555.
\newline\noindent
The simulations used in this work were performed on the DiRAC@Durham facility managed by the Institute for Computational Cosmology on behalf of the STFC DiRAC HPC Facility (www.dirac.ac.uk). The equipment was funded by BEIS capital funding via STFC capital grants ST/P002293/1, ST/R002371/1 and ST/S002502/1, Durham University and STFC operations grant ST/R000832/1. DiRAC is part of the National e-Infrastructure.
The post-processing analysis for this project utilized the UCloud interactive HPC system managed by the eScience Center at the University of Southern Denmark. 
\newline\newline
{\bf Author contribution statement:} SMK introduced redshift drift computations into \texttt{mescaline} under heavy guidance from HJM. HJM introduced the LTB model and redshift drift computations along radial lines of sight in the LTB model as a test for \texttt{mescaline} under guidance from SMK. AH made analytical assessments necessary for these efforts.
Data and presented figures were produced by SMK and HJM. 
All authors contributed significantly to debugging and project development as well as to the writing of the manuscript. 
%
\end{acknowledgments}

\appendix

\section{Precision and accuracy test using the Lema\^itre-Tolman-Bondi metric}\label{app:LTBtest} 

In this appendix we test our implementation of redshift drift computations in \texttt{mescaline} by verifying that the computations can reproduce known results for the Lema\^itre-Tolman-Bondi (LTB) metric.
The LTB models are spherically symmetric dust solutions to Einstein's equations which may include a non-vanishing cosmological constant. 
The line element of the LTB models can be written in spherical coordinates as
\begin{align}
    ds^2 = -dt^2 + R(t,r)dr^2 + A^2(t,r)d\Omega^2,
\end{align}
where $R(t,r) := A_{,r}^2(t,r)/\left[1-k(r)\right]$, $k(r)$ is the spatial curvature, and $d\Omega\equiv d\theta^2 + \sin^2(\theta)d\phi^2$. Ignoring a possible cosmological constant, the metric function $A(t,r)$ fulfills the dynamical equation
\begin{align}\label{eq:dAdt}
    A^2_{,t}(t,r) = \frac{2M(r)}{A(t,r)} - k(r) \, ,
\end{align} 
where $M(r)$ is a constant of integration in the time coordinate and is related to the dust rest-frame density according to 
\begin{align}
    \rho(t,r) = \frac{M_{,r}(r)}{4\pi A^2 A_{,r}}.
\end{align}
To specify a particular LTB model we adopt the following curvature function
\begin{equation}\label{eq:kr}
k(r) =
    \begin{cases}
        -k_{\rm max}r^2\left[\left(\frac{r}{r_b} \right) ^n-1 \right] ^m & \text{if } r\leq r_b \\
        0 & \text{if } r > r_b
    \end{cases}
\end{equation}
with different values of $n,m$ (either equal to 4 or 6) and with varying values of $k_{\rm max}$ around the order of $10^{-8}$, and varying values of $r_b$. To uniquely specify the models we set $A(t_0, r) = a_{\rm EdS}(t_0)\cdot r$, where $a_{\rm EdS}$ is the EdS scale factor with reduced Hubble parameter $h = 0.45$. 
To find the function $M(r)$ which corresponds to our chosen model, we will use  
the relation \citep{Bolejko_2005}
\begin{align}
 t - t_B =  \frac{M}{(-k)^{3/2}} \bigg[ &\sqrt{\bigg(1-\frac{kA}{M}\bigg)^2-1} \\
 & - {\rm cosh}^{-1}\bigg(1-\frac{kA}{M}\bigg) \bigg],\nonumber
\end{align} 
where all functional dependence on the coordinates $t$ and $r$ is from now on left implicit in the equations.
The function $t_B$ is the $r-$dependent time of the Big Bang in the model, here set to vanish identically. 
To solve the above equation, given our initial function $A(t_0,r)$, $k(r)$, and $t_B=0$, we use a Newton-Raphson root finding method to find the $M$ which satisfies this relation at each value of $r$ to within a tolerance of $10^{-8}$ times the typical size of terms in the equation. This fully specifies the metric functions for the LTB model, which can then be integrated back in time alongside the \texttt{mescaline} ray tracer. To minimise finite-difference derivatives, we evolve $A_{,r}(t,r)$ alongside $A(t,r)$ using the following equation
\begin{align}
	A_{,tr} &= \left( \frac{M_{,r}}{AA_{,t}} - \frac{MA_{,r}}{A^2A_{,t}} - \frac{k_{,r}}{2A_{,t}} \right)\label{eq:dAdtdr}.
\end{align}
We additionally make use of the above relation and 
\begin{align}
	A_{,ttr} &= \left(-\frac{M_{,r}}{A^2} + \frac{2MA_{,r}}{A^3} \right)\label{eq:dAdt2dr}
\end{align}
to obtain $A_{,tr}$ and $A_{,ttr}$ without having to resort to finite differences.

To advance the LTB metric we specify initial data for $A(t_0,r)$ as above and choose values of $r_b, k_{\rm max}$ and the exponents $n,m$ to specify the curvature $k(r)$ (which is constant in time). These functions are set up on a Cartesian cubic grid using the coordinate transforms given in Appendix~\ref{app:transforms} below. We then use these functions, alongside our root finder, to converge to a function $M(r)$ at each point on the grid. All functions are set to zero at the origin $r=0$. This forms our complete initial data for the LTB metric. Next, we advance the system of equations \eqref{eq:dAdt}, \eqref{eq:dAdtdr}, and \eqref{eq:dAdt2dr} using a Runge-Kutta $4^{\rm th}$ order (RK-4) integration scheme backwards in time from redshift $z=0$. The LTB metric functions are used to fill the components of the metric and extrinsic curvature tensors, the latter having non-vanishing components
\begin{align}
	K_{rr} &= -\frac{A_{,r}A_{,tr}}{1-k}\\
	K_{\theta\theta} &= -AA_{,t}\\
	K_{\phi\phi} &= -AA_{,t}\sin^2(\theta).
\end{align}
Once computed in spherical coordinates, these are transformed to the Cartesian grid (see Appendix~\ref{app:transforms}). The metric on the cubic grid is sent into the test version of the \texttt{mescaline} ray tracer \citep[see][]{Macpherson:2023}, including the redshift drift calculation discussed in the main text. 

We wish to test our redshift drift calculation compared to the analytic LTB redshift drift. For radial light rays in LTB models, the redshift drift can be computed by solving the following coupled set of equations \citep{LTB2}
\begin{align}\label{eqs:dzLTB_ana}
    \begin{split}
    \frac{dz}{dr} &= (1+z)\frac{A_{,tr}}{\sqrt{1-k}},\\
    \frac{dt}{dr}&= -\frac{A_{,r}}{\sqrt{1-k}},\\
    \frac{d\delta z}{dr}& = \frac{A_{,tr}\delta z}{\sqrt{1-k}} + (1+z)\frac{A_{,ttr}\delta t}{\sqrt{1-k}},\\
    \frac{d\delta t}{dr}& = -\frac{A_{,tr}\delta t}{\sqrt{1-k}},
    \end{split}
\end{align}
where we have enforced $r$ of the observer to be smaller than $r$ of the emitter in choosing sign conventions. We choose initial data for our semi-analytic solution as $z=\delta z=0$ and $\delta t=30$ years and evolve the equations above using an RK-4 integrator. The precision of the main redshift drift calculations is first order (since we approximate the integral using a sum), so we can treat our semi-analytic redshift drift computations as the `true' analytic solutions since they have far subdominant error for the same resolution. 

For all test cases here, we place the observer several grid cells from the origin at $r=0$. This is to avoid the origin being used in finite-difference derivatives for quantities involved in our calculation. Since the metric is spherically-symmetric, we shoot only one light ray for this test and use only one observer. We test three specific choices of the size of the LTB structure, namely: $r_b=50,700$, and 1000 $h^{-1}$ Mpc. For all cases we use $k_{\rm max}=9.9\times 10^{-8}$ and $n=4 = m$ except for the case with $r_b = 1000h^{-1}$Mpc where we use a steeper density profile by setting $n = 6= m$.
We change the size of the structure primarily to adjust the amount of time the light ray spends inside and outside the LTB structure before it reaches the observer.
We perform tests for three Cartesian grid resolutions of $N=32, 64$, and 128, each with a cubic box length of $L=1\,h^{-1}$ Gpc. In each case, we calculate the redshift drift using \texttt{mescaline} with the LTB metric, $\delta z_{\rm num}$, and compare to the semi-analytic from advancing \eqref{eqs:dzLTB_ana}, $\delta z_{\rm ana}$, to get the error at resolution $i$: $\delta_i \equiv \delta z_{\rm num} / \delta z_{\rm ana} - 1$. We assess the convergence of the error when increasing resolution by calculating the convergence factor $(\delta_1-\delta_2)/(\delta_2 - \delta_3)$ where resolution indices $i=1,2,3$ represent grid resolutions of $N=32,63,128$, respectively. For our first-order approximation of the sum for the redshift drift, the rate of convergence should be equal to 2 \citep[see also Appendix~B of][]{Adamek:2020}. We note that in each resolution, the observer is placed at a slightly different location. For each case, we place the observer $\sim10$ grid cells from the origin at $r=0$ to ensure our finite-difference stencils do not ever overlap the $r=0$ point.

\begin{figure}
    \centering
    \includegraphics[scale = 0.5]{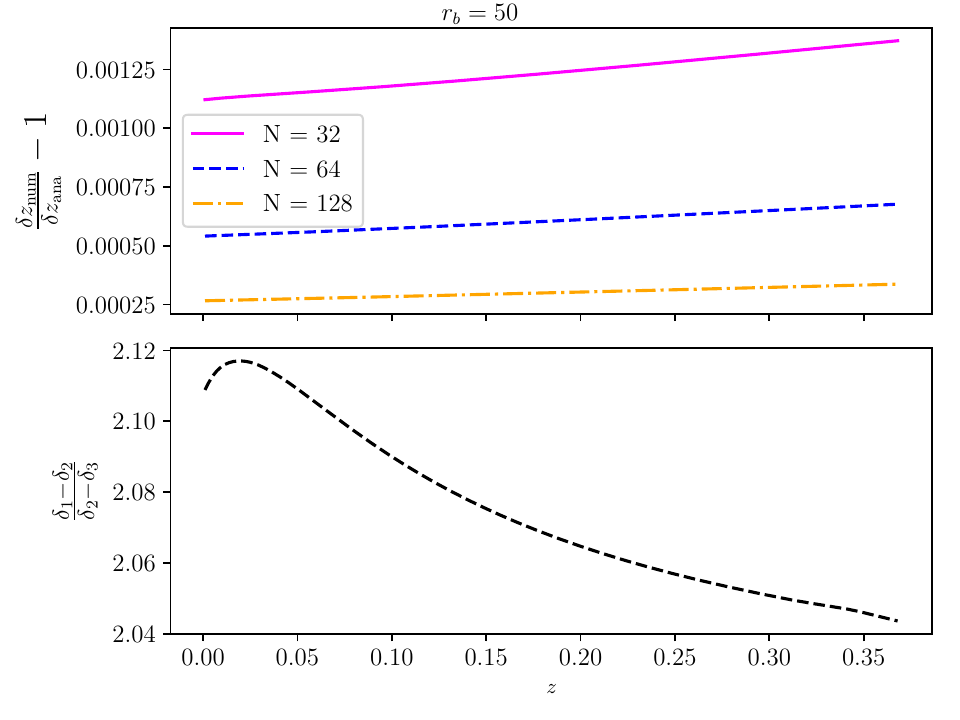}
    \caption{Top panel: Redshift drift error for the LTB model with $r_b=50\,h^{-1}$ Mpc and $m=n=4$ in $k(r)$ for three resolutions $N=32,64,$ and 128. Bottom panel: convergence factor for the three lines in the top panel. The expected convergence rate is 2.}
    \label{fig:C_rb50}
\end{figure}
Figure~\ref{fig:C_rb50} shows the redshift drift error as a function of redshift for three resolutions $N=32, 64$, and 128 (solid, dashed, and dot-dashed curves in the top panel, respectively) for an LTB structure with $r_b=50\,h^{-1}$ Mpc and $m=n=4$ in \eqref{eq:kr}. In this case, since the observer is slightly off-centre, the structure is so small that the observer actually lies in a pure FLRW region. This represents a sanity check of our code that the metric is accurately being set to the FLRW region in this case. In the bottom panel, the convergence factor is always $\approx 2$, as expected for our first-order scheme. 
 
\begin{figure}
    \centering
    \includegraphics[scale = 0.5]{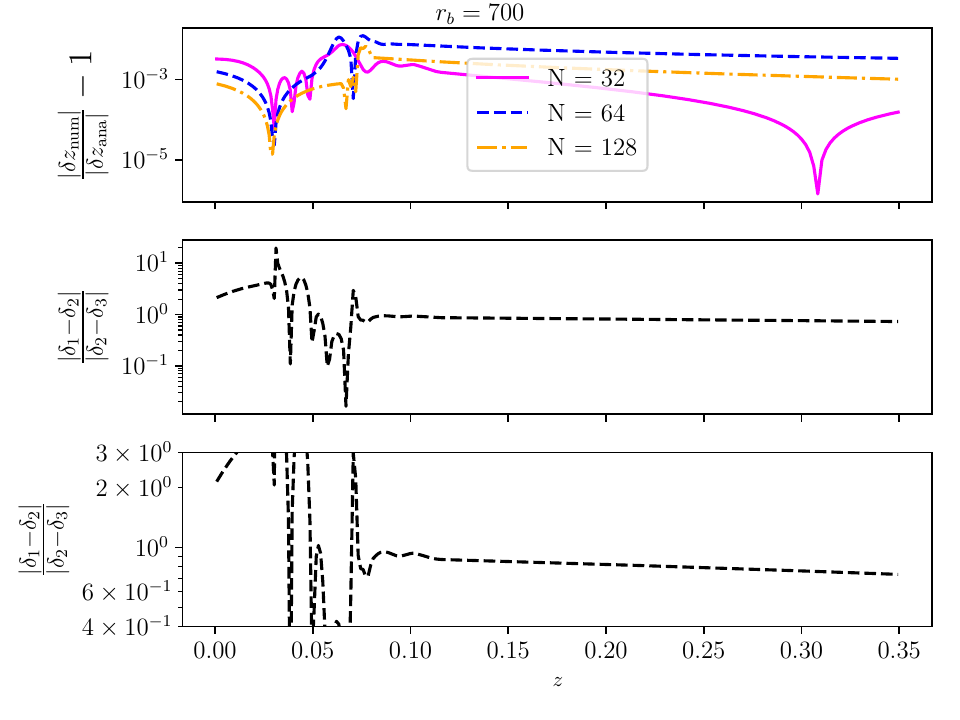}
    \caption{Top panel: Redshift drift error for the LTB model with $r_b=700\,h^{-1}$ Mpc and $m=n=4$ in $k(r)$ for three resolutions $N=32,64,$ and 128. Middle and bottom panel: convergence factor for the three lines in the top panel (the bottom panel shows a zoomed-in version of the middle panel). The expected convergence rate is 2.}
    \label{fig:C_rb700}
\end{figure}
The top panel of Figure~\ref{fig:C_rb700} shows the redshift drift error for an LTB structure with $r_b=700\,h^{-1}$ Mpc and $m=n=4$ in \eqref{eq:kr}. In this case, the observer is inside the LTB structure but quite close to the edge, such that there is only a little propagation inside the structure before the ray moves into the FLRW region. The middle panel shows the convergence rate and the bottom panel shows a zoomed-in $y$-axis of the middle panel. 

The LTB functions we set up are effectively discontinuous at the edge of the structure, due to the step function we define for $k(r)$. Such a function affects the calculation of finite-difference derivatives---an important part of the redshift drift calculation to get the Ricci and Weyl tensors---when they cross over such a boundary. At $z<0.03$, we see the expected convergence rate of 2. However, between $z\approx$0.04--0.06 we see the convergence is spoiled due to the overlapping of our finite-difference stencil with the boundary of the LTB structure. After the ray passes outside the structure into the FLRW region, the convergence is still spoiled. Since the redshift drift is calculated as a sum over the line of sight, any error introduced earlier in the calculation (i.e., at lower redshift) will propagate to the rest of the calculation (i.e., higher redshift). In this case, we see some convergence from $N=64$ to $N=128$, though not for $N=32$. A potential explanation for this could be because the observer is quite close to the LTB structure, the contribution from the early steps of the calculation (and thus, the error in crossing the boundary), is subdominant with respect to the redshift drift signal at $z>0.1$. However, it could be that the error for the case of $N=32$ is sufficient in this case to spoil the convergence even at higher $z$. 

\begin{figure}
    \centering
    \includegraphics[scale = 0.5]{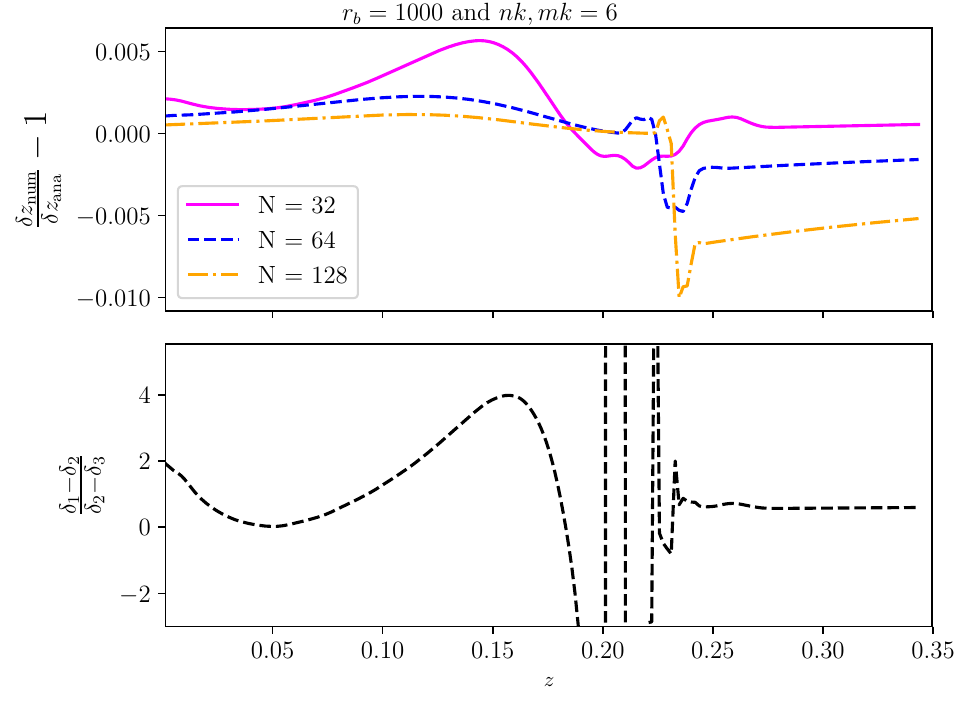}
    \caption{Top panel: Redshift drift error for the LTB model with $r_b=1000\,h^{-1}$ Mpc and $m=n=6$ in $k(r)$ for three resolutions $N=32,64,$ and 128. Bottom panel: convergence factor for the three lines in the top panel. The expected convergence rate is 2.}
    \label{fig:C_rb1000}
\end{figure}
Figure~\ref{fig:C_rb1000} shows the error in the redshift drift (top panel) and convergence rate (bottom panel) for the LTB metric with $r_b=1000\,h^{-1}$ Mpc and $m=n=6$ in \eqref{eq:kr}. In this case the observer is quite far from the LTB structure boundary so a large portion of the rays propagation is inside the structure itself. In this case, we see good convergence for the first few steps which is quickly spoiled by the $N=32$ test case. We see good convergence for $N=64$ to $N=128$ until our finite-difference derivatives overlap the LTB structure at $z\approx$0.15--0.2. The convergence is also spoiled after the ray has passed through the structure boundary. 
The non-convergence of the $N=32$ test inside the LTB structure is most likely due to the fact that this is simply too low a numerical resolution to accurately resolve the steep gradients of this structure. The fact that the $N=64$ and $N=128$ errors are more well-behaved supports this explanation. We expect that if a third, higher resolution run was included in place of the $N=32$ test that in this final case the convergence would be equal to 2 inside the LTB structure before the boundary is reached. 

In this appendix, we have shown that the redshift drift calculation we present in the main text matches the analytic LTB solution when expected. Our results are satisfactory in terms of their convergence rate in most cases when the LTB structure boundary does not influence the calculation. Since this boundary is artificial, we will not encounter these kinds of issues with our calculations in the simulations presented in the main text. Of importance is also the magnitude of the relative error in the redshift drift: in all cases with the highest resolution we use here ($N=128$) the error is less than $\sim 0.02$--0.1\% depending on the structure we test. We expect the error on our redshift drift calculations in the ET simulations to be even below this value for two reasons. Firstly, we use an even higher resolution of $N=256$ for the calculations in the main text. Secondly, the LTB structure we test here is extreme in its density contrasts. This implies that derivatives of metric functions will be less accurately approximated than if the structure had lower density fluctuations. Our ET simulations do not contain structures as extreme as the LTB model we study here, so we expect our finite-difference derivatives to be better approximations of metric derivatives. 
However, in the case of our simulations we have an additional source of error from the evolution of the metric itself. In the next section, we perform a rough convergence test to show that this error is subdominant with respect to our results.

\section{Convergence of simulation results}\label{app:128}
In the previous appendix, we isolated the error on our redshift drift calculations with \texttt{mescaline} and showed that we find satisfactory errors which converge with resolution as expected in the case of an LTB test metric. 

We would like to quantify the numerical errors for the main simulations of this paper.
However, the resolution of the simulations set a scale of truncation below which structures cannot form.
In our case, due to the non-linear nature of our simulations, an increase in resolution will allow structures to form at scales below the initial resolution scale, making a strict Richardson extrapolation test invalid for placing error bars on our calculations. 
This also makes the identification of ``the same observer'' across different resolutions of the simulations ill defined.

In place of such a test, in this appendix we ensure that our simulations have \textit{statistically} converged. 
Concretely, we want to ensure that the values of the redshift drift measurements across different lines of sights and synthetic observers are robust in a distributional sense (mean and variance) towards an increase in resolution.
We thus perform a lower-resolution simulation ($N=128$) which is statistically similar to the $N=256$ simulation presented in the main text. The higher-resolution simulation contains more small-scale structure than the lower-resolution simulation. However, on the scales where we are sampling the redshift drift ($z\approx 0.5$) the structures should be statistically similar. 
We perform the redshift drift calculation for a set of observers in each simulation (10 observers with $N_{\rm side}=32$ for the $N=128$ simulation and 50 observers with $N_{\rm side}=8$ in the $N=256$ simulation) and compare our main results: the mean and variance of the redshift drift over all lines of sight and the separate contributions from the Ricci and Weyl components of the signal. Potential differences between the calculations in the two simulations can be attributed to a) differences in numerical resolution (i.e., an increased truncation error in the $N=128$ simulation) and/or b) differences in physical structure (i.e., an increased amplitude of small-scale fluctuations in the $N=256$ simulation). When integrating over the intermediate redshifts we study here ($z\approx 0.5$--1) to calculate the redshift drift, we expect b) to be subdominant/average out \citep[due to the fact that similar simulations contain little backreaction on these scales; see][]{Macpherson:2019a}. 
Thus, we expect the main source of differences between the simulations to be due to a). However, it is important to note on small scales (low redshifts), that b) can become important. 
 
\begin{figure}
    \centering
    \includegraphics[scale = 0.5]{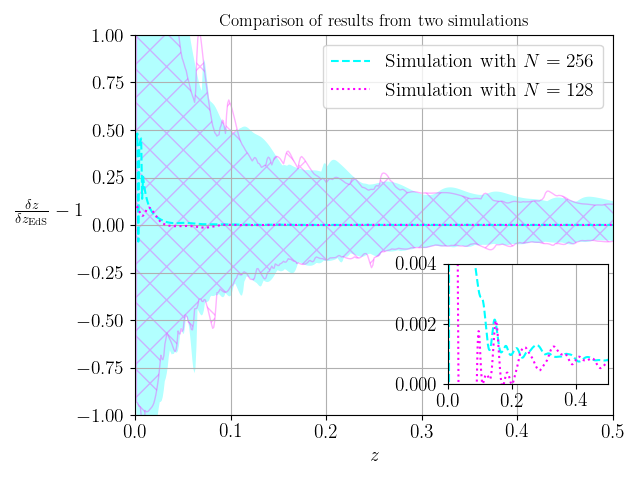}
    \caption{Mean redshift drift for the $N=128$ (dotted curve) and the $N=256$ (dashed  curve) resolution simulations as a function of redshift. The variance from the maximum to minimum fluctuation is shown as a shaded region for $N=256$ and a hatched region for $N=128$. The inset shows a zoomed-in version of the $y$-axis. The redshift drift mean and fluctuations are shown relative to the EdS prediction. 
    }
    \label{fig:dz_128}
\end{figure}
Figure~\ref{fig:dz_128} shows a comparison of the mean (curves) and variance (curves and shaded areas) of the redshift drift in the two simulations. The redshift drift is shown relative to the EdS prediction and the variance is shown as 68.1\% and 95.4\% intervals. Comparing the mean values for both simulations as a function of redshift (inset in Figure~\ref{fig:dz_128}), we can see that the low-redshift fluctuations in the mean redshift drift are not robust to the resolution changes in this test. 
The most likely cause for this is the increased amplitude of small-scale fluctuations in the $N=256$ simulation (mostly due to additional modes below the minimum resolution scale of the $N=128$ simulation). However, already at $z\approx 0.1$, the difference in mean has dropped to below 1\%. 
Overall, we should thus be wary in attributing physical meaning to the fluctuation visible in the mean redshift drift below $z\approx 0.2$. 
Beyond this redshift, to the maximum $z=0.5$ studied here (due to the smaller box size of the $N=128$ simulation), the mean redshift drift agrees between the two simulations. We thus conclude that our main results for $z\gtrsim 0.2$ are robust to changes in resolution. 

The fluctuations in the redshift drift are robust to the resolution changes in this test for $z\gtrsim 0.02$. 
Thus, our main text results of the fluctuations in the redshift drift should also only be considered meaningful beyond this redshift.

\begin{figure*}
    \centering
    \subfigure[]{\includegraphics[scale = 0.5]{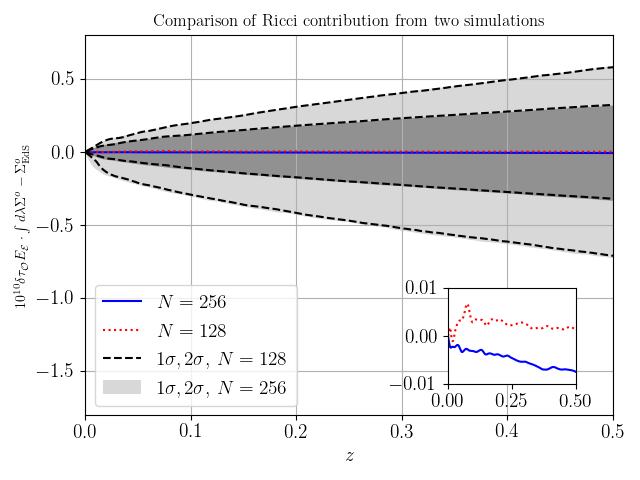}}
    \subfigure[]{\includegraphics[scale = 0.5]{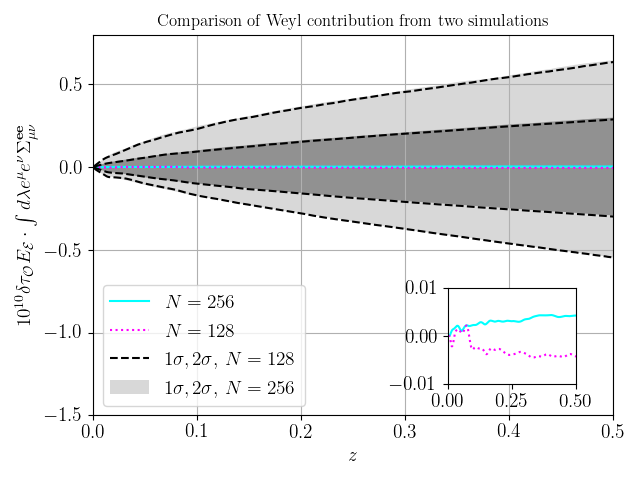}}
    \caption{Mean and fluctuations of the Ricci ($\delta \tau_\obs E_{\mathcal{E}}\int d\lambda \Sigma^o-\Sigma^o_{\rm EdS}$; left panel) and Weyl ($\delta \tau_\obs E_{\mathcal{E}} \int d\lambda e^\mu e^\nu \Sigma^{\bf ee}_{\mu\nu}$; right panel) contributions to the redshift drift. Mean values are shown as curves, with a zoom-in of the y-axis shown in the inset to emphasize the difference in mean values. The 68.1\% and 95.4\% confidence intervals are shown as dark and light grey shaded regions for $N=256$ and dashed black curves for $N=128$ in both panels. 
    }
    \label{fig:WeyldRicci_128}
\end{figure*}
In Figure~\ref{fig:WeyldRicci_128} we compare the Weyl and Ricci contributions to the redshift drift for the two simulations. 
The mean values for each term are very close to zero for both resolutions. We see these mean values change sign (yet maintain the same order of magnitude amplitude of $\sim$0.5--1\%) with a change in resolution. 
This could be explained simply by our relatively low number of lines of sight in this study when comparing resolutions. Specifically, the expected error from finite number of lines of sight is $1/\sqrt{N_{\rm LOS}}\approx 0.005$ here, which is the same amplitude as the mean values of the Ricci and Weyl contributions in the inset of Figure~\ref{fig:WeyldRicci_128}.
Further, the mean value of these contributions could be dominated by numerical error from the simulations due to their small amplitude. However, the cancellation of the two contributions is robust to resolution changes (resulting in the consistent mean redshift drift we find in Figure~\ref{fig:dz_128}). 

In Figure~\ref{fig:WeyldRicci_128} the 68.1\% and 95.4\% confidence intervals across all light rays for the Ricci (left panel) and Weyl (right panel) components are indicated with dark and light grey shaded regions for $N=256$ and black dashed curves for $N=128$. 
A general trend we see here is that the Ricci fluctuations are skewed towards negative values and the Weyl fluctuations are skewed towards positive values. This quality is robust to the changes in resolution we see here, as well as the order of magnitude of the fluctuations we find. 
Of note is the fact that the maximum of the Ricci contribution and the minimum of the Weyl contribution is consistent between the simulations. However, the minimum of the Ricci and the maximum of the Weyl---corresponding to the light ray with the largest absolute value for each case---changes by a factor of up to $\sim 1.5$ with changes in resolution. This difference is most likely to be due to sample variance since we find that the confidence intervals are very similar for the two simulations. The differences could also be partially attributed to the increased numerical error in the $N=128$ simulation.
 
Due to the nature of this rough convergence test, we are not able to fully determine the importance of these possibilities. 
An important point in this test is the different numbers of observers and lines of sight from the two simulations. The lower-resolution $N=128$ simulation contains $\sim 3$ times more total lines of sight over all observers, in addition to a larger total number of observers---implying more independent lines of sight. This difference could contribute to the differences we see in Figure~\ref{fig:WeyldRicci_128}. We could potentially test the significance of light ray statistics on our results by comparing equal numbers of light rays and observers between the simulations. Due to the computational expense of adding more observers to the $N=128$ simulation, this would require analysing a subset of observers for $N=256$ and down-grading the number of lines of sight for the 10 observers in the $N=128$ simulation. In this case, we would be studying only a total of 7680 lines of sight for each simulation, which may not be sufficient statistics for a meaningful comparison.

For now, we conclude that the qualitative aspect of the overall order-of-magnitude of the separate Ricci and Weyl fluctuations (and not their mean values), as well as the skewness of both distributions, is robust to changes in resolution. We leave a thorough investigation into the precise amplitude of these individual effects to future work.

\section{Spherical to Cartesian coordinate transforms}\label{app:transforms}

Since the LTB metric is spherically symmetric about the origin, it is easier to describe and initialise an LTB model using spherical coordinates. 
\texttt{Mescaline} assumes a cubic grid in Cartesian coordinates as input, so to introduce the LTB spacetime into \texttt{mescaline}, we need to make a transformation from spherical to Cartesian coordinates. For this we use the relations
\begin{align}
    \begin{split}
    x &= r\sin(\theta)\cos(\phi),	\\
    y&= r\sin(\theta)\sin(\phi),\\
    z &= r\cos(\theta),\\
    r &= \sqrt{x^2+y^2+z^2},\\
    \theta & = 2\cdot \rm atan\left(\frac{\sqrt{x^2+y^2}}{\sqrt{x^2+y^2+z^2}+z} \right) = \rm atan2\left(\sqrt{x^2+y^2},z \right),  \\
    \phi & = 2\cdot \rm atan\left(\frac{y}{\sqrt{x^2+y^2}+x} \right) = \rm atan2(y,x).
    \end{split}
\end{align}
We then use the standard tensor transformation rule, e.g., $g_{\mu\nu} = \frac{\partial x^{\tilde \mu}}{\partial x^\mu}\frac{\partial x^{\tilde \nu}}{\partial x^\nu}g_{\tilde \mu\tilde\nu}$. The partial derivatives of the coordinates are:
\begin{align}\label{eq:partials}
    \begin{split}
    \frac{\partial x}{\partial r} & = \sin(\theta)\cos(\phi) = \frac{x}{r} = \frac{\partial r}{\partial x}\\
    \frac{\partial y}{\partial r} & = \sin(\theta)\sin(\phi) = \frac{y}{r} = \frac{\partial r}{\partial y}\\
    \frac{\partial z}{\partial r} & = \cos(\theta) = \frac{z}{r} = \frac{\partial r}{\partial z}\\
    \frac{\partial \theta}{\partial x} & = \frac{x}{r^2}\frac{\cos(\theta)}{\sin(\theta)}\\
    \frac{\partial \phi}{\partial x} & = -\left( \frac{y}{r^2}+\frac{1}{r}\frac{\cos^2(\theta)}{\sin(\theta)}\sin(\phi) \right) \\ &= -y\frac{ \sqrt{x^2+y^2} +x }{ (x^2+y^2)^{3/2} + x^3 + y^2x} \\
    \frac{\partial \theta}{\partial y}  & = \frac{y}{r^2}\frac{\cos(\theta)}{\sin(\theta)} = \frac{\sin(\phi)\cos(\theta)}{r}\\
    \frac{\partial \phi}{\partial y} & = \frac{x}{r^2} + \frac{1}{r}\frac{\cos^2(\theta)}{\sin(\theta)}\cos(\phi) =\\ & x\frac{\sqrt{x^2+y^2}+x}{(x^2+y^2)^{3/2}+xy^2+x^3}\\
    \frac{\partial \theta}{\partial z} & = \frac{1}{r\sin(\theta)}\left( \frac{z}{r}\cos(\theta)-1 \right) \\= &\frac{\cos(\phi)}{x}\left( \frac{z}{r}\cos(\theta)-1 \right) = \frac{\sin(\phi)}{y}\left( \frac{z}{r}\cos(\theta)-1 \right)  \\
    \frac{\partial x}{\partial \theta} & = r\cos(\theta)\cos(\phi)\\
    \frac{\partial x}{\partial \phi} & = -r\sin(\theta)\sin(\phi) = -y\\
    \frac{\partial y}{\partial \theta} & = r\cos(\theta)\sin(\phi)\\
    \frac{\partial y}{\partial \phi} & = r\sin(\theta)\cos(\phi) = x\\
    \frac{\partial z}{\partial \theta} & -r\sin(\theta)\\
    \frac{\partial z}{\partial \phi} & = 0 =     \frac{\partial \phi}{\partial z}.
    \end{split}
\end{align}

\section{Power spectrum convergence test}\label{app:power_spectrum}
\begin{figure*}

\centering
\includegraphics[scale = 0.5]{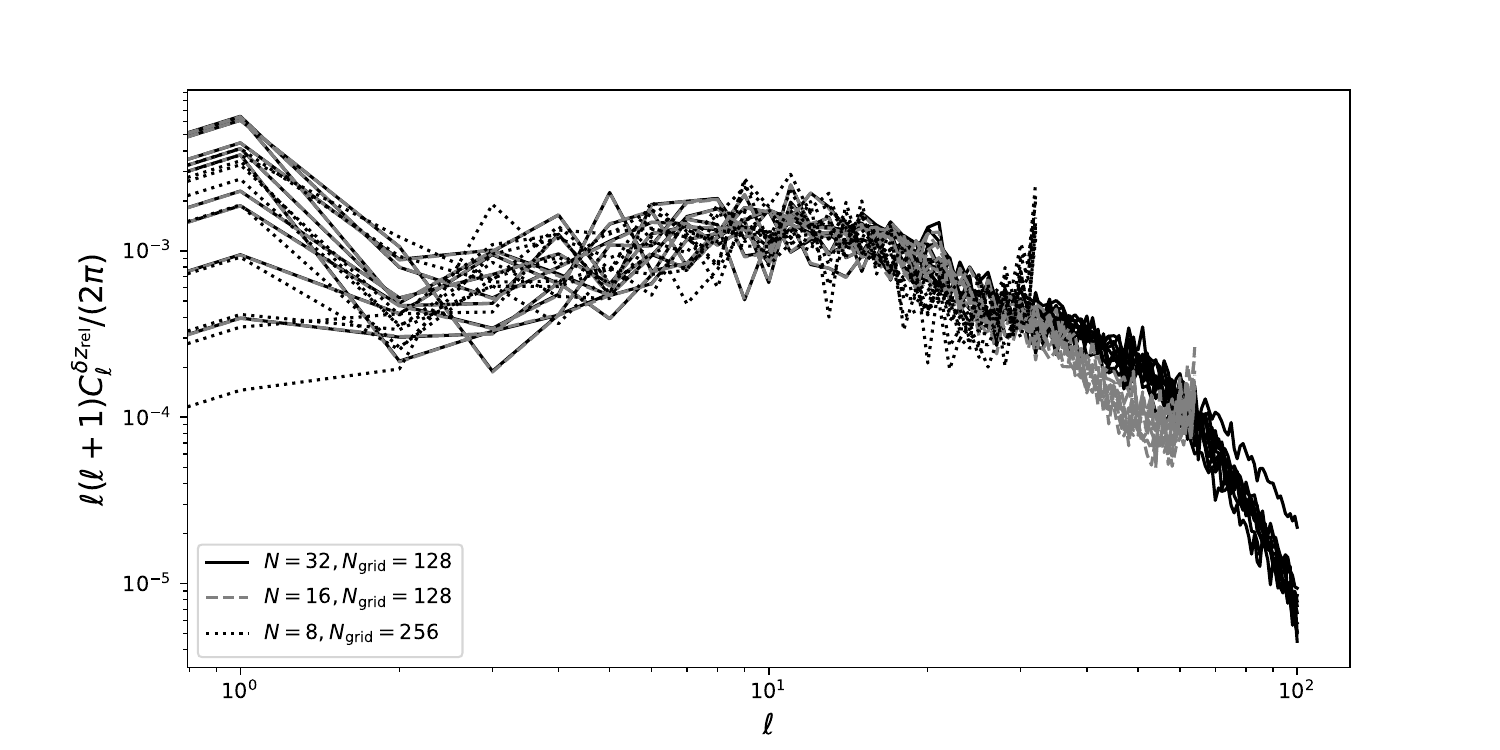}
\caption{Angular power spectra of redshift drift skymaps for 10 observers for each of three combinations of $N$ and $N_{\rm grid}.$
}
\label{fig:pp_test}
\end{figure*}

In this appendix, we present a convergence test of the redshift drift power spectra in order to verify that the results presented in figure \ref{fig:dz_Cls} are not dominated by noise. Figure \ref{fig:pp_test} compares the power spectra for 10 observers for each of three combinations of $N$ and $N_{\rm side}$. The first combination corresponds to the results shown in figure \ref{fig:dz_Cls} and thus correspond to $N = 128$ and $N_{\rm side} = 32$. We remind that $N_{\rm side} = 32$ means that each observer has $12\cdot 32^2$ lines of sight and that the maximum value of $\ell$ in the corresponding power spectrum is $4\cdot N_{\rm side}$. The second combination of $N$ and $N_{\rm side}$ are $N = 128$ and $N_{\rm side} = 16$ while the third combination is $N = 256$ and $N_{\rm side} = 8$. All power spectra are for $z \approx 0.1$. The power spectra are shown in figure \ref{fig:pp_test}. The power spectra overall agree very well for all three combinations of $N$ and $N_{\rm side}$. We therefore conclude that the power spectra in figure \ref{fig:dz_Cls} are not dominated by noise.

\clearpage
\bibliography{refs}

\begin{thebibliography}{62}%
\makeatletter
\providecommand \@ifxundefined [1]{%
 \@ifx{#1\undefined}
}%
\providecommand \@ifnum [1]{%
 \ifnum #1\expandafter \@firstoftwo
 \else \expandafter \@secondoftwo
 \fi
}%
\providecommand \@ifx [1]{%
 \ifx #1\expandafter \@firstoftwo
 \else \expandafter \@secondoftwo
 \fi
}%
\providecommand \natexlab [1]{#1}%
\providecommand \enquote  [1]{``#1''}%
\providecommand \bibnamefont  [1]{#1}%
\providecommand \bibfnamefont [1]{#1}%
\providecommand \citenamefont [1]{#1}%
\providecommand \href@noop [0]{\@secondoftwo}%
\providecommand \href [0]{\begingroup \@sanitize@url \@href}%
\providecommand \@href[1]{\@@startlink{#1}\@@href}%
\providecommand \@@href[1]{\endgroup#1\@@endlink}%
\providecommand \@sanitize@url [0]{\catcode `\\12\catcode `\$12\catcode
  `\&12\catcode `\#12\catcode `\^12\catcode `\_12\catcode `\%12\relax}%
\providecommand \@@startlink[1]{}%
\providecommand \@@endlink[0]{}%
\providecommand \url  [0]{\begingroup\@sanitize@url \@url }%
\providecommand \@url [1]{\endgroup\@href {#1}{\urlprefix }}%
\providecommand \urlprefix  [0]{URL }%
\providecommand \Eprint [0]{\href }%
\providecommand \doibase [0]{https://doi.org/}%
\providecommand \selectlanguage [0]{\@gobble}%
\providecommand \bibinfo  [0]{\@secondoftwo}%
\providecommand \bibfield  [0]{\@secondoftwo}%
\providecommand \translation [1]{[#1]}%
\providecommand \BibitemOpen [0]{}%
\providecommand \bibitemStop [0]{}%
\providecommand \bibitemNoStop [0]{.\EOS\space}%
\providecommand \EOS [0]{\spacefactor3000\relax}%
\providecommand \BibitemShut  [1]{\csname bibitem#1\endcsname}%
\let\auto@bib@innerbib\@empty
\bibitem [{\citenamefont {Sandage}(1962)}]{Sandage}%
  \BibitemOpen
  \bibfield  {author} {\bibinfo {author} {\bibfnamefont {A.~R.}\ \bibnamefont
  {Sandage}},\ }\bibfield  {title} {\bibinfo {title} {The change of redshift
  and apparent luminosity of galaxies due to the deceleration of selected
  expanding universes.},\ }\href
  {https://api.semanticscholar.org/CorpusID:119959450} {\bibfield  {journal}
  {\bibinfo  {journal} {The Astrophysical Journal}\ }\textbf {\bibinfo {volume}
  {136}},\ \bibinfo {pages} {319} (\bibinfo {year} {1962})}\BibitemShut
  {NoStop}%
\bibitem [{\citenamefont {Mcvittie}(1962)}]{Mcvittie}%
  \BibitemOpen
  \bibfield  {author} {\bibinfo {author} {\bibfnamefont {G.~C.}\ \bibnamefont
  {Mcvittie}},\ }\bibfield  {title} {\bibinfo {title} {Appendix to the change
  of redshift and apparent luminosity of galaxies due to the deceleration of
  selected expanding universes.},\ }\href
  {https://api.semanticscholar.org/CorpusID:118736788} {\bibfield  {journal}
  {\bibinfo  {journal} {The Astrophysical Journal}\ }\textbf {\bibinfo {volume}
  {136}},\ \bibinfo {pages} {334} (\bibinfo {year} {1962})}\BibitemShut
  {NoStop}%
\bibitem [{\citenamefont {Klöckner}\ \emph {et~al.}(2015)\citenamefont
  {Klöckner}, \citenamefont {Obreschkow}, \citenamefont {Martins},
  \citenamefont {Raccanelli}, \citenamefont {Champion}, \citenamefont {Roy},
  \citenamefont {Lobanov}, \citenamefont {Wagner},\ and\ \citenamefont
  {Keller}}]{SKA}%
  \BibitemOpen
  \bibfield  {author} {\bibinfo {author} {\bibfnamefont {H.~R.}\ \bibnamefont
  {Klöckner}}, \bibinfo {author} {\bibfnamefont {D.}~\bibnamefont
  {Obreschkow}}, \bibinfo {author} {\bibfnamefont {C.}~\bibnamefont {Martins}},
  \bibinfo {author} {\bibfnamefont {A.}~\bibnamefont {Raccanelli}}, \bibinfo
  {author} {\bibfnamefont {D.}~\bibnamefont {Champion}}, \bibinfo {author}
  {\bibfnamefont {A.}~\bibnamefont {Roy}}, \bibinfo {author} {\bibfnamefont
  {A.}~\bibnamefont {Lobanov}}, \bibinfo {author} {\bibfnamefont
  {J.}~\bibnamefont {Wagner}},\ and\ \bibinfo {author} {\bibfnamefont
  {R.}~\bibnamefont {Keller}},\ }\href@noop {} {\bibinfo {title} {Real time
  cosmology - a direct measure of the expansion rate of the universe}}
  (\bibinfo {year} {2015}),\ \Eprint {https://arxiv.org/abs/1501.03822}
  {arXiv:1501.03822 [astro-ph.CO]} \BibitemShut {NoStop}%
\bibitem [{\citenamefont {Rocha}\ and\ \citenamefont {Martins}(2022)}]{SKA2}%
  \BibitemOpen
  \bibfield  {author} {\bibinfo {author} {\bibfnamefont {B.~A.~R.}\
  \bibnamefont {Rocha}}\ and\ \bibinfo {author} {\bibfnamefont {C.~J. A.~P.}\
  \bibnamefont {Martins}},\ }\bibfield  {title} {\bibinfo {title} {Redshift
  drift cosmography with elt and skao measurements},\ }\href
  {https://doi.org/10.1093/mnras/stac3240} {\bibfield  {journal} {\bibinfo
  {journal} {Monthly Notices of the Royal Astronomical Society}\ }\textbf
  {\bibinfo {volume} {518}},\ \bibinfo {pages} {2853–2869} (\bibinfo {year}
  {2022})}\BibitemShut {NoStop}%
\bibitem [{\citenamefont {Yoo}\ \emph {et~al.}(2011)\citenamefont {Yoo},
  \citenamefont {Kai},\ and\ \citenamefont {Nakao}}]{2011PhRvD..83d3527Y}%
  \BibitemOpen
  \bibfield  {author} {\bibinfo {author} {\bibfnamefont {C.-M.}\ \bibnamefont
  {Yoo}}, \bibinfo {author} {\bibfnamefont {T.}~\bibnamefont {Kai}},\ and\
  \bibinfo {author} {\bibfnamefont {K.-i.}\ \bibnamefont {Nakao}},\ }\bibfield
  {title} {\bibinfo {title} {Redshift drift in lemaître-tolman-bondi void
  universes},\ }\bibfield  {journal} {\bibinfo  {journal} {Physical Review D}\
  }\textbf {\bibinfo {volume} {83}},\ \href
  {https://doi.org/10.1103/physrevd.83.043527} {10.1103/physrevd.83.043527}
  (\bibinfo {year} {2011})\BibitemShut {NoStop}%
\bibitem [{\citenamefont {Codur}\ and\ \citenamefont {Marinoni}(2021)}]{LTB2}%
  \BibitemOpen
  \bibfield  {author} {\bibinfo {author} {\bibfnamefont {R.}~\bibnamefont
  {Codur}}\ and\ \bibinfo {author} {\bibfnamefont {C.}~\bibnamefont
  {Marinoni}},\ }\href@noop {} {\bibinfo {title} {Redshift drift in radially
  inhomogeneous lemaitre-tolman-bondi spacetimes}} (\bibinfo {year} {2021}),\
  \Eprint {https://arxiv.org/abs/2107.04868} {arXiv:2107.04868 [gr-qc]}
  \BibitemShut {NoStop}%
\bibitem [{\citenamefont {Balcerzak}\ and\ \citenamefont
  {Dabrowski}(2013)}]{stephani}%
  \BibitemOpen
  \bibfield  {author} {\bibinfo {author} {\bibfnamefont {A.}~\bibnamefont
  {Balcerzak}}\ and\ \bibinfo {author} {\bibfnamefont {M.~P.}\ \bibnamefont
  {Dabrowski}},\ }\bibfield  {title} {\bibinfo {title} {Redshift drift in a
  pressure-gradient cosmology},\ }\bibfield  {journal} {\bibinfo  {journal}
  {Physical Review D}\ }\textbf {\bibinfo {volume} {87}},\ \href
  {https://doi.org/10.1103/physrevd.87.063506} {10.1103/physrevd.87.063506}
  (\bibinfo {year} {2013})\BibitemShut {NoStop}%
\bibitem [{\citenamefont {Mishra}\ \emph
  {et~al.}(2012{\natexlab{a}})\citenamefont {Mishra}, \citenamefont
  {Célérier},\ and\ \citenamefont {Singh}}]{Szekeres1}%
  \BibitemOpen
  \bibfield  {author} {\bibinfo {author} {\bibfnamefont {P.}~\bibnamefont
  {Mishra}}, \bibinfo {author} {\bibfnamefont {M.-N.}\ \bibnamefont
  {Célérier}},\ and\ \bibinfo {author} {\bibfnamefont {T.~P.}\ \bibnamefont
  {Singh}},\ }\bibfield  {title} {\bibinfo {title} {Redshift drift in axially
  symmetric quasispherical szekeres models},\ }\bibfield  {journal} {\bibinfo
  {journal} {Physical Review D}\ }\textbf {\bibinfo {volume} {86}},\ \href
  {https://doi.org/10.1103/physrevd.86.083520} {10.1103/physrevd.86.083520}
  (\bibinfo {year} {2012}{\natexlab{a}})\BibitemShut {NoStop}%
\bibitem [{\citenamefont {Mishra}\ and\ \citenamefont
  {Célérier}(2014)}]{Szekeres2}%
  \BibitemOpen
  \bibfield  {author} {\bibinfo {author} {\bibfnamefont {P.}~\bibnamefont
  {Mishra}}\ and\ \bibinfo {author} {\bibfnamefont {M.-N.}\ \bibnamefont
  {Célérier}},\ }\href@noop {} {\bibinfo {title} {Redshift and redshift-drift
  in $\lambda = 0$ quasi-spherical szekeres cosmological models and the effect
  of averaging}} (\bibinfo {year} {2014}),\ \Eprint
  {https://arxiv.org/abs/1403.5229} {arXiv:1403.5229 [astro-ph.CO]}
  \BibitemShut {NoStop}%
\bibitem [{\citenamefont {Koksbang}\ and\ \citenamefont
  {Heinesen}(2022)}]{Koksbang_2022}%
  \BibitemOpen
  \bibfield  {author} {\bibinfo {author} {\bibfnamefont {S.~M.}\ \bibnamefont
  {Koksbang}}\ and\ \bibinfo {author} {\bibfnamefont {A.}~\bibnamefont
  {Heinesen}},\ }\bibfield  {title} {\bibinfo {title} {Redshift drift in a
  universe with structure: Lemaitre-tolman-bondi structures with arbitrary
  angle of entry of light},\ }\bibfield  {journal} {\bibinfo  {journal}
  {Physical Review D}\ }\textbf {\bibinfo {volume} {106}},\ \href
  {https://doi.org/10.1103/physrevd.106.043501} {10.1103/physrevd.106.043501}
  (\bibinfo {year} {2022})\BibitemShut {NoStop}%
\bibitem [{\citenamefont {Koksbang}\ and\ \citenamefont
  {Hannestad}(2016)}]{Koksbang:2015ctu}%
  \BibitemOpen
  \bibfield  {author} {\bibinfo {author} {\bibfnamefont {S.~M.}\ \bibnamefont
  {Koksbang}}\ and\ \bibinfo {author} {\bibfnamefont {S.}~\bibnamefont
  {Hannestad}},\ }\bibfield  {title} {\bibinfo {title} {{Redshift drift in an
  inhomogeneous universe: averaging and the backreaction conjecture}},\ }\href
  {https://doi.org/10.1088/1475-7516/2016/01/009} {\bibfield  {journal}
  {\bibinfo  {journal} {JCAP}\ }\textbf {\bibinfo {volume} {01}},\ \bibinfo
  {pages} {009}},\ \Eprint {https://arxiv.org/abs/1512.05624} {arXiv:1512.05624
  [astro-ph.CO]} \BibitemShut {NoStop}%
\bibitem [{\citenamefont {Wiltshire}(2009)}]{Wiltshire_2009}%
  \BibitemOpen
  \bibfield  {author} {\bibinfo {author} {\bibfnamefont {D.~L.}\ \bibnamefont
  {Wiltshire}},\ }\bibfield  {title} {\bibinfo {title} {Average observational
  quantities in the timescape cosmology},\ }\bibfield  {journal} {\bibinfo
  {journal} {Physical Review D}\ }\textbf {\bibinfo {volume} {80}},\ \href
  {https://doi.org/10.1103/physrevd.80.123512} {10.1103/physrevd.80.123512}
  (\bibinfo {year} {2009})\BibitemShut {NoStop}%
\bibitem [{\citenamefont {Fleury}\ \emph {et~al.}(2015)\citenamefont {Fleury},
  \citenamefont {Pitrou},\ and\ \citenamefont {Uzan}}]{Bianchi1}%
  \BibitemOpen
  \bibfield  {author} {\bibinfo {author} {\bibfnamefont {P.}~\bibnamefont
  {Fleury}}, \bibinfo {author} {\bibfnamefont {C.}~\bibnamefont {Pitrou}},\
  and\ \bibinfo {author} {\bibfnamefont {J.-P.}\ \bibnamefont {Uzan}},\
  }\bibfield  {title} {\bibinfo {title} {Light propagation in a homogeneous and
  anisotropic universe},\ }\bibfield  {journal} {\bibinfo  {journal} {Physical
  Review D}\ }\textbf {\bibinfo {volume} {91}},\ \href
  {https://doi.org/10.1103/physrevd.91.043511} {10.1103/physrevd.91.043511}
  (\bibinfo {year} {2015})\BibitemShut {NoStop}%
\bibitem [{\citenamefont {Marcori}\ \emph {et~al.}(2018)\citenamefont
  {Marcori}, \citenamefont {Pitrou}, \citenamefont {Uzan},\ and\ \citenamefont
  {Pereira}}]{Bianchi2}%
  \BibitemOpen
  \bibfield  {author} {\bibinfo {author} {\bibfnamefont {O.~H.}\ \bibnamefont
  {Marcori}}, \bibinfo {author} {\bibfnamefont {C.}~\bibnamefont {Pitrou}},
  \bibinfo {author} {\bibfnamefont {J.-P.}\ \bibnamefont {Uzan}},\ and\
  \bibinfo {author} {\bibfnamefont {T.~S.}\ \bibnamefont {Pereira}},\
  }\bibfield  {title} {\bibinfo {title} {Direction and redshift drifts for
  general observers and their applications in cosmology},\ }\bibfield
  {journal} {\bibinfo  {journal} {Physical Review D}\ }\textbf {\bibinfo
  {volume} {98}},\ \href {https://doi.org/10.1103/physrevd.98.023517}
  {10.1103/physrevd.98.023517} (\bibinfo {year} {2018})\BibitemShut {NoStop}%
\bibitem [{\citenamefont {Migkas}\ \emph {et~al.}(2020)\citenamefont {Migkas},
  \citenamefont {Schellenberger}, \citenamefont {Reiprich}, \citenamefont
  {Pacaud}, \citenamefont {Ramos-Ceja},\ and\ \citenamefont
  {Lovisari}}]{anisotropy2}%
  \BibitemOpen
  \bibfield  {author} {\bibinfo {author} {\bibfnamefont {K.}~\bibnamefont
  {Migkas}}, \bibinfo {author} {\bibfnamefont {G.}~\bibnamefont
  {Schellenberger}}, \bibinfo {author} {\bibfnamefont {T.~H.}\ \bibnamefont
  {Reiprich}}, \bibinfo {author} {\bibfnamefont {F.}~\bibnamefont {Pacaud}},
  \bibinfo {author} {\bibfnamefont {M.~E.}\ \bibnamefont {Ramos-Ceja}},\ and\
  \bibinfo {author} {\bibfnamefont {L.}~\bibnamefont {Lovisari}},\ }\bibfield
  {title} {\bibinfo {title} {Probing cosmic isotropy with a new x-ray galaxy
  cluster sample through thelx–tscaling relation},\ }\href
  {https://doi.org/10.1051/0004-6361/201936602} {\bibfield  {journal} {\bibinfo
   {journal} {Astronomy \&; Astrophysics}\ }\textbf {\bibinfo {volume} {636}},\
  \bibinfo {pages} {A15} (\bibinfo {year} {2020})}\BibitemShut {NoStop}%
\bibitem [{\citenamefont {Migkas}\ \emph {et~al.}(2021)\citenamefont {Migkas},
  \citenamefont {Pacaud}, \citenamefont {Schellenberger}, \citenamefont
  {Erler}, \citenamefont {Nguyen-Dang}, \citenamefont {Reiprich}, \citenamefont
  {Ramos-Ceja},\ and\ \citenamefont {Lovisari}}]{anisotropy3}%
  \BibitemOpen
  \bibfield  {author} {\bibinfo {author} {\bibfnamefont {K.}~\bibnamefont
  {Migkas}}, \bibinfo {author} {\bibfnamefont {F.}~\bibnamefont {Pacaud}},
  \bibinfo {author} {\bibfnamefont {G.}~\bibnamefont {Schellenberger}},
  \bibinfo {author} {\bibfnamefont {J.}~\bibnamefont {Erler}}, \bibinfo
  {author} {\bibfnamefont {N.~T.}\ \bibnamefont {Nguyen-Dang}}, \bibinfo
  {author} {\bibfnamefont {T.~H.}\ \bibnamefont {Reiprich}}, \bibinfo {author}
  {\bibfnamefont {M.~E.}\ \bibnamefont {Ramos-Ceja}},\ and\ \bibinfo {author}
  {\bibfnamefont {L.}~\bibnamefont {Lovisari}},\ }\bibfield  {title} {\bibinfo
  {title} {Cosmological implications of the anisotropy of ten galaxy cluster
  scaling relations},\ }\href {https://doi.org/10.1051/0004-6361/202140296}
  {\bibfield  {journal} {\bibinfo  {journal} {Astronomy \&; Astrophysics}\
  }\textbf {\bibinfo {volume} {649}},\ \bibinfo {pages} {A151} (\bibinfo {year}
  {2021})}\BibitemShut {NoStop}%
\bibitem [{\citenamefont {Blake}\ and\ \citenamefont
  {Wall}(2002)}]{anisotropy4}%
  \BibitemOpen
  \bibfield  {author} {\bibinfo {author} {\bibfnamefont {C.}~\bibnamefont
  {Blake}}\ and\ \bibinfo {author} {\bibfnamefont {J.}~\bibnamefont {Wall}},\
  }\bibfield  {title} {\bibinfo {title} {A velocity dipole in the distribution
  of radio galaxies},\ }\href {https://doi.org/10.1038/416150a} {\bibfield
  {journal} {\bibinfo  {journal} {Nature}\ }\textbf {\bibinfo {volume} {416}},\
  \bibinfo {pages} {150–152} (\bibinfo {year} {2002})}\BibitemShut {NoStop}%
\bibitem [{\citenamefont {Secrest}\ \emph {et~al.}(2021)\citenamefont
  {Secrest}, \citenamefont {Hausegger}, \citenamefont {Rameez}, \citenamefont
  {Mohayaee}, \citenamefont {Sarkar},\ and\ \citenamefont
  {Colin}}]{anisotropy5}%
  \BibitemOpen
  \bibfield  {author} {\bibinfo {author} {\bibfnamefont {N.~J.}\ \bibnamefont
  {Secrest}}, \bibinfo {author} {\bibfnamefont {S.~v.}\ \bibnamefont
  {Hausegger}}, \bibinfo {author} {\bibfnamefont {M.}~\bibnamefont {Rameez}},
  \bibinfo {author} {\bibfnamefont {R.}~\bibnamefont {Mohayaee}}, \bibinfo
  {author} {\bibfnamefont {S.}~\bibnamefont {Sarkar}},\ and\ \bibinfo {author}
  {\bibfnamefont {J.}~\bibnamefont {Colin}},\ }\bibfield  {title} {\bibinfo
  {title} {A test of the cosmological principle with quasars},\ }\href
  {https://doi.org/10.3847/2041-8213/abdd40} {\bibfield  {journal} {\bibinfo
  {journal} {The Astrophysical Journal Letters}\ }\textbf {\bibinfo {volume}
  {908}},\ \bibinfo {pages} {L51} (\bibinfo {year} {2021})}\BibitemShut
  {NoStop}%
\bibitem [{\citenamefont {Secrest}\ \emph {et~al.}(2022)\citenamefont
  {Secrest}, \citenamefont {von Hausegger}, \citenamefont {Rameez},
  \citenamefont {Mohayaee},\ and\ \citenamefont {Sarkar}}]{anisotropy6}%
  \BibitemOpen
  \bibfield  {author} {\bibinfo {author} {\bibfnamefont {N.~J.}\ \bibnamefont
  {Secrest}}, \bibinfo {author} {\bibfnamefont {S.}~\bibnamefont {von
  Hausegger}}, \bibinfo {author} {\bibfnamefont {M.}~\bibnamefont {Rameez}},
  \bibinfo {author} {\bibfnamefont {R.}~\bibnamefont {Mohayaee}},\ and\
  \bibinfo {author} {\bibfnamefont {S.}~\bibnamefont {Sarkar}},\ }\bibfield
  {title} {\bibinfo {title} {A challenge to the standard cosmological model},\
  }\href {https://doi.org/10.3847/2041-8213/ac88c0} {\bibfield  {journal}
  {\bibinfo  {journal} {The Astrophysical Journal Letters}\ }\textbf {\bibinfo
  {volume} {937}},\ \bibinfo {pages} {L31} (\bibinfo {year}
  {2022})}\BibitemShut {NoStop}%
\bibitem [{\citenamefont {Dam}\ \emph {et~al.}(2023)\citenamefont {Dam},
  \citenamefont {Lewis},\ and\ \citenamefont {Brewer}}]{anisotropy7}%
  \BibitemOpen
  \bibfield  {author} {\bibinfo {author} {\bibfnamefont {L.}~\bibnamefont
  {Dam}}, \bibinfo {author} {\bibfnamefont {G.~F.}\ \bibnamefont {Lewis}},\
  and\ \bibinfo {author} {\bibfnamefont {B.~J.}\ \bibnamefont {Brewer}},\
  }\bibfield  {title} {\bibinfo {title} {Testing the cosmological principle
  with catwise quasars: a bayesian analysis of the number-count dipole},\
  }\href {https://doi.org/10.1093/mnras/stad2322} {\bibfield  {journal}
  {\bibinfo  {journal} {Monthly Notices of the Royal Astronomical Society}\
  }\textbf {\bibinfo {volume} {525}},\ \bibinfo {pages} {231–245} (\bibinfo
  {year} {2023})}\BibitemShut {NoStop}%
\bibitem [{\citenamefont {Colin}\ \emph {et~al.}(2019)\citenamefont {Colin},
  \citenamefont {Mohayaee}, \citenamefont {Rameez},\ and\ \citenamefont
  {Sarkar}}]{anisotropy8}%
  \BibitemOpen
  \bibfield  {author} {\bibinfo {author} {\bibfnamefont {J.}~\bibnamefont
  {Colin}}, \bibinfo {author} {\bibfnamefont {R.}~\bibnamefont {Mohayaee}},
  \bibinfo {author} {\bibfnamefont {M.}~\bibnamefont {Rameez}},\ and\ \bibinfo
  {author} {\bibfnamefont {S.}~\bibnamefont {Sarkar}},\ }\bibfield  {title}
  {\bibinfo {title} {Evidence for anisotropy of cosmic acceleration},\ }\href
  {https://doi.org/10.1051/0004-6361/201936373} {\bibfield  {journal} {\bibinfo
   {journal} {Astronomy \&; Astrophysics}\ }\textbf {\bibinfo {volume} {631}},\
  \bibinfo {pages} {L13} (\bibinfo {year} {2019})}\BibitemShut {NoStop}%
\bibitem [{\citenamefont {{Ferreira}}\ and\ \citenamefont
  {{Quartin}}(2021)}]{Ferreira:2021wr}%
  \BibitemOpen
  \bibfield  {author} {\bibinfo {author} {\bibfnamefont {P.~d.~S.}\
  \bibnamefont {{Ferreira}}}\ and\ \bibinfo {author} {\bibfnamefont
  {M.}~\bibnamefont {{Quartin}}},\ }\bibfield  {title} {\bibinfo {title}
  {{First Constraints on the Intrinsic CMB Dipole and Our Velocity with Doppler
  and Aberration}},\ }\href {https://doi.org/10.1103/PhysRevLett.127.101301}
  {\bibfield  {journal} {\bibinfo  {journal} {\prl}\ }\textbf {\bibinfo
  {volume} {127}},\ \bibinfo {eid} {101301} (\bibinfo {year}
  {2021})}\BibitemShut {NoStop}%
\bibitem [{\citenamefont {{Darling}}(2022)}]{Darling:2022ue}%
  \BibitemOpen
  \bibfield  {author} {\bibinfo {author} {\bibfnamefont {J.}~\bibnamefont
  {{Darling}}},\ }\bibfield  {title} {\bibinfo {title} {{The Universe is
  Brighter in the Direction of Our Motion: Galaxy Counts and Fluxes are
  Consistent with the CMB Dipole}},\ }\href
  {https://doi.org/10.3847/2041-8213/ac6f08} {\bibfield  {journal} {\bibinfo
  {journal} {apjl}\ }\textbf {\bibinfo {volume} {931}},\ \bibinfo {eid} {L14}
  (\bibinfo {year} {2022})},\ \Eprint {https://arxiv.org/abs/2205.06880}
  {arXiv:2205.06880 [astro-ph.CO]} \BibitemShut {NoStop}%
\bibitem [{\citenamefont {{Horstmann}}\ \emph {et~al.}(2022)\citenamefont
  {{Horstmann}}, \citenamefont {{Pietschke}},\ and\ \citenamefont
  {{Schwarz}}}]{Horstmann:2022}%
  \BibitemOpen
  \bibfield  {author} {\bibinfo {author} {\bibfnamefont {N.}~\bibnamefont
  {{Horstmann}}}, \bibinfo {author} {\bibfnamefont {Y.}~\bibnamefont
  {{Pietschke}}},\ and\ \bibinfo {author} {\bibfnamefont {D.~J.}\ \bibnamefont
  {{Schwarz}}},\ }\bibfield  {title} {\bibinfo {title} {{Inference of the
  cosmic rest-frame from supernovae Ia}},\ }\href
  {https://doi.org/10.1051/0004-6361/202142640} {\bibfield  {journal} {\bibinfo
   {journal} {aap}\ }\textbf {\bibinfo {volume} {668}},\ \bibinfo {eid} {A34}
  (\bibinfo {year} {2022})},\ \Eprint {https://arxiv.org/abs/2111.03055}
  {arXiv:2111.03055 [astro-ph.CO]} \BibitemShut {NoStop}%
\bibitem [{\citenamefont {{Akarsu}}\ \emph {et~al.}(2023)\citenamefont
  {{Akarsu}}, \citenamefont {{Di Valentino}}, \citenamefont {{Kumar}},
  \citenamefont {{{\"O}zyi{\u{g}}it}},\ and\ \citenamefont
  {{Sharma}}}]{Akarsu:2023ud}%
  \BibitemOpen
  \bibfield  {author} {\bibinfo {author} {\bibfnamefont {{\"O}.}~\bibnamefont
  {{Akarsu}}}, \bibinfo {author} {\bibfnamefont {E.}~\bibnamefont {{Di
  Valentino}}}, \bibinfo {author} {\bibfnamefont {S.}~\bibnamefont {{Kumar}}},
  \bibinfo {author} {\bibfnamefont {M.}~\bibnamefont {{{\"O}zyi{\u{g}}it}}},\
  and\ \bibinfo {author} {\bibfnamefont {S.}~\bibnamefont {{Sharma}}},\
  }\bibfield  {title} {\bibinfo {title} {{Testing spatial curvature and
  anisotropic expansion on top of the {\ensuremath{\Lambda}}CDM model}},\
  }\href {https://doi.org/10.1016/j.dark.2022.101162} {\bibfield  {journal}
  {\bibinfo  {journal} {Physics of the Dark Universe}\ }\textbf {\bibinfo
  {volume} {39}},\ \bibinfo {eid} {101162} (\bibinfo {year} {2023})},\ \Eprint
  {https://arxiv.org/abs/2112.07807} {arXiv:2112.07807 [astro-ph.CO]}
  \BibitemShut {NoStop}%
\bibitem [{\citenamefont {Kumar~Aluri}\ \emph {et~al.}(2023)\citenamefont
  {Kumar~Aluri}, \citenamefont {Cea}, \citenamefont {Chingangbam},
  \citenamefont {Chu}, \citenamefont {Clowes}, \citenamefont {Hutsemékers},
  \citenamefont {Kochappan}, \citenamefont {Lopez}, \citenamefont {Liu},
  \citenamefont {Martens}, \citenamefont {Martins}, \citenamefont {Migkas},
  \citenamefont {Ó~Colgáin}, \citenamefont {Pranav}, \citenamefont {Shamir},
  \citenamefont {Singal}, \citenamefont {Sheikh-Jabbari}, \citenamefont
  {Wagner}, \citenamefont {Wang}, \citenamefont {Wiltshire}, \citenamefont
  {Yeung}, \citenamefont {Yin},\ and\ \citenamefont {Zhao}}]{anisotropy1}%
  \BibitemOpen
  \bibfield  {author} {\bibinfo {author} {\bibfnamefont {P.}~\bibnamefont
  {Kumar~Aluri}}, \bibinfo {author} {\bibfnamefont {P.}~\bibnamefont {Cea}},
  \bibinfo {author} {\bibfnamefont {P.}~\bibnamefont {Chingangbam}}, \bibinfo
  {author} {\bibfnamefont {M.-C.}\ \bibnamefont {Chu}}, \bibinfo {author}
  {\bibfnamefont {R.~G.}\ \bibnamefont {Clowes}}, \bibinfo {author}
  {\bibfnamefont {D.}~\bibnamefont {Hutsemékers}}, \bibinfo {author}
  {\bibfnamefont {J.~P.}\ \bibnamefont {Kochappan}}, \bibinfo {author}
  {\bibfnamefont {A.~M.}\ \bibnamefont {Lopez}}, \bibinfo {author}
  {\bibfnamefont {L.}~\bibnamefont {Liu}}, \bibinfo {author} {\bibfnamefont
  {N.~C.~M.}\ \bibnamefont {Martens}}, \bibinfo {author} {\bibfnamefont {C.~J.
  A.~P.}\ \bibnamefont {Martins}}, \bibinfo {author} {\bibfnamefont
  {K.}~\bibnamefont {Migkas}}, \bibinfo {author} {\bibfnamefont
  {E.}~\bibnamefont {Ó~Colgáin}}, \bibinfo {author} {\bibfnamefont
  {P.}~\bibnamefont {Pranav}}, \bibinfo {author} {\bibfnamefont
  {L.}~\bibnamefont {Shamir}}, \bibinfo {author} {\bibfnamefont {A.~K.}\
  \bibnamefont {Singal}}, \bibinfo {author} {\bibfnamefont {M.~M.}\
  \bibnamefont {Sheikh-Jabbari}}, \bibinfo {author} {\bibfnamefont
  {J.}~\bibnamefont {Wagner}}, \bibinfo {author} {\bibfnamefont {S.-J.}\
  \bibnamefont {Wang}}, \bibinfo {author} {\bibfnamefont {D.~L.}\ \bibnamefont
  {Wiltshire}}, \bibinfo {author} {\bibfnamefont {S.}~\bibnamefont {Yeung}},
  \bibinfo {author} {\bibfnamefont {L.}~\bibnamefont {Yin}},\ and\ \bibinfo
  {author} {\bibfnamefont {W.}~\bibnamefont {Zhao}},\ }\bibfield  {title}
  {\bibinfo {title} {Is the observable universe consistent with the
  cosmological principle?},\ }\href {https://doi.org/10.1088/1361-6382/acbefc}
  {\bibfield  {journal} {\bibinfo  {journal} {Classical and Quantum Gravity}\
  }\textbf {\bibinfo {volume} {40}},\ \bibinfo {pages} {094001} (\bibinfo
  {year} {2023})}\BibitemShut {NoStop}%
\bibitem [{\citenamefont {Bessa}\ \emph {et~al.}(2023)\citenamefont {Bessa},
  \citenamefont {Durrer},\ and\ \citenamefont {Stock}}]{pert_durrer}%
  \BibitemOpen
  \bibfield  {author} {\bibinfo {author} {\bibfnamefont {P.}~\bibnamefont
  {Bessa}}, \bibinfo {author} {\bibfnamefont {R.}~\bibnamefont {Durrer}},\ and\
  \bibinfo {author} {\bibfnamefont {D.}~\bibnamefont {Stock}},\ }\bibfield
  {title} {\bibinfo {title} {Perturbations of cosmological redshift drift},\
  }\href {https://doi.org/10.1088/1475-7516/2023/11/093} {\bibfield  {journal}
  {\bibinfo  {journal} {Journal of Cosmology and Astroparticle Physics}\
  }\textbf {\bibinfo {volume} {2023}}\bibinfo  {number} { (11)},\ \bibinfo
  {pages} {093}}\BibitemShut {NoStop}%
\bibitem [{\citenamefont {Kim}\ \emph {et~al.}(2015)\citenamefont {Kim},
  \citenamefont {Linder}, \citenamefont {Edelstein},\ and\ \citenamefont
  {Erskine}}]{pert1}%
  \BibitemOpen
\bibfield  {number} {  }\bibfield  {author} {\bibinfo {author} {\bibfnamefont
  {A.~G.}\ \bibnamefont {Kim}}, \bibinfo {author} {\bibfnamefont {E.~V.}\
  \bibnamefont {Linder}}, \bibinfo {author} {\bibfnamefont {J.}~\bibnamefont
  {Edelstein}},\ and\ \bibinfo {author} {\bibfnamefont {D.}~\bibnamefont
  {Erskine}},\ }\bibfield  {title} {\bibinfo {title} {Giving cosmic redshift
  drift a whirl},\ }\href {https://doi.org/10.1016/j.astropartphys.2014.09.004}
  {\bibfield  {journal} {\bibinfo  {journal} {Astroparticle Physics}\ }\textbf
  {\bibinfo {volume} {62}},\ \bibinfo {pages} {195–205} (\bibinfo {year}
  {2015})}\BibitemShut {NoStop}%
\bibitem [{\citenamefont {Linder}(2010)}]{pert2}%
  \BibitemOpen
  \bibfield  {author} {\bibinfo {author} {\bibfnamefont {E.~V.}\ \bibnamefont
  {Linder}},\ }\href@noop {} {\bibinfo {title} {Constraining models of dark
  energy}} (\bibinfo {year} {2010}),\ \Eprint {https://arxiv.org/abs/1004.4646}
  {arXiv:1004.4646 [astro-ph.CO]} \BibitemShut {NoStop}%
\bibitem [{\citenamefont {Koksbang}(2023)}]{Koksbang_2023}%
  \BibitemOpen
  \bibfield  {author} {\bibinfo {author} {\bibfnamefont {S.~M.}\ \bibnamefont
  {Koksbang}},\ }\bibfield  {title} {\bibinfo {title} {Redshift drift in a
  universe with structure. ii. light rays propagated through a newtonian n-body
  simulation},\ }\bibfield  {journal} {\bibinfo  {journal} {Physical Review D}\
  }\textbf {\bibinfo {volume} {107}},\ \href
  {https://doi.org/10.1103/physrevd.107.063544} {10.1103/physrevd.107.063544}
  (\bibinfo {year} {2023})\BibitemShut {NoStop}%
\bibitem [{\citenamefont {Korzy\'nski}\ and\ \citenamefont
  {Kopi\'nski}(2018)}]{Korzynski:2017nas}%
  \BibitemOpen
  \bibfield  {author} {\bibinfo {author} {\bibfnamefont {M.}~\bibnamefont
  {Korzy\'nski}}\ and\ \bibinfo {author} {\bibfnamefont {J.}~\bibnamefont
  {Kopi\'nski}},\ }\bibfield  {title} {\bibinfo {title} {{Optical drift effects
  in general relativity}},\ }\href
  {https://doi.org/10.1088/1475-7516/2018/03/012} {\bibfield  {journal}
  {\bibinfo  {journal} {JCAP}\ }\textbf {\bibinfo {volume} {03}},\ \bibinfo
  {pages} {012}},\ \Eprint {https://arxiv.org/abs/1711.00584} {arXiv:1711.00584
  [gr-qc]} \BibitemShut {NoStop}%
\bibitem [{\citenamefont {Heinesen}(2021{\natexlab{a}})}]{Heinesen:2020pms}%
  \BibitemOpen
  \bibfield  {author} {\bibinfo {author} {\bibfnamefont {A.}~\bibnamefont
  {Heinesen}},\ }\bibfield  {title} {\bibinfo {title} {{Multipole decomposition
  of redshift drift -- model independent mapping of the expansion history of
  the Universe}},\ }\href {https://doi.org/10.1103/PhysRevD.103.023537}
  {\bibfield  {journal} {\bibinfo  {journal} {Phys. Rev. D}\ }\textbf {\bibinfo
  {volume} {103}},\ \bibinfo {pages} {023537} (\bibinfo {year}
  {2021}{\natexlab{a}})},\ \Eprint {https://arxiv.org/abs/2011.10048}
  {arXiv:2011.10048 [gr-qc]} \BibitemShut {NoStop}%
\bibitem [{\citenamefont {Heinesen}(2021{\natexlab{b}})}]{Heinesen:2021nrc}%
  \BibitemOpen
  \bibfield  {author} {\bibinfo {author} {\bibfnamefont {A.}~\bibnamefont
  {Heinesen}},\ }\bibfield  {title} {\bibinfo {title} {{Redshift drift as a
  model independent probe of dark energy}},\ }\href
  {https://doi.org/10.1103/PhysRevD.103.L081302} {\bibfield  {journal}
  {\bibinfo  {journal} {Phys. Rev. D}\ }\textbf {\bibinfo {volume} {103}},\
  \bibinfo {pages} {L081302} (\bibinfo {year} {2021}{\natexlab{b}})},\ \Eprint
  {https://arxiv.org/abs/2102.03774} {arXiv:2102.03774 [gr-qc]} \BibitemShut
  {NoStop}%
\bibitem [{\citenamefont {Heinesen}(2021{\natexlab{c}})}]{Heinesen:2021qnl}%
  \BibitemOpen
  \bibfield  {author} {\bibinfo {author} {\bibfnamefont {A.}~\bibnamefont
  {Heinesen}},\ }\bibfield  {title} {\bibinfo {title} {{Redshift drift
  cosmography for model-independent cosmological inference}},\ }\href
  {https://doi.org/10.1103/PhysRevD.104.123527} {\bibfield  {journal} {\bibinfo
   {journal} {Phys. Rev. D}\ }\textbf {\bibinfo {volume} {104}},\ \bibinfo
  {pages} {123527} (\bibinfo {year} {2021}{\natexlab{c}})},\ \Eprint
  {https://arxiv.org/abs/2107.08674} {arXiv:2107.08674 [astro-ph.CO]}
  \BibitemShut {NoStop}%
\bibitem [{\citenamefont {Koksbang}(2019)}]{Koksbang:2019glb}%
  \BibitemOpen
  \bibfield  {author} {\bibinfo {author} {\bibfnamefont {S.~M.}\ \bibnamefont
  {Koksbang}},\ }\bibfield  {title} {\bibinfo {title} {{Another look at
  redshift drift and the backreaction conjecture}},\ }\href
  {https://doi.org/10.1088/1475-7516/2019/10/036} {\bibfield  {journal}
  {\bibinfo  {journal} {JCAP}\ }\textbf {\bibinfo {volume} {10}},\ \bibinfo
  {pages} {036}},\ \Eprint {https://arxiv.org/abs/1909.13489} {arXiv:1909.13489
  [astro-ph.CO]} \BibitemShut {NoStop}%
\bibitem [{\citenamefont {Koksbang}(2020)}]{Koksbang:2020zej}%
  \BibitemOpen
  \bibfield  {author} {\bibinfo {author} {\bibfnamefont {S.~M.}\ \bibnamefont
  {Koksbang}},\ }\bibfield  {title} {\bibinfo {title} {{Observations in
  statistically homogeneous, locally inhomogeneous cosmological toy-models
  without FLRW backgrounds}},\ }\href {https://doi.org/10.1093/mnrasl/slaa146}
  {\bibfield  {journal} {\bibinfo  {journal} {Mon. Not. Roy. Astron. Soc.}\
  }\textbf {\bibinfo {volume} {498}},\ \bibinfo {pages} {L135} (\bibinfo {year}
  {2020})},\ \bibinfo {note} {[Erratum: Mon.Not.Roy.Astron.Soc. 500, (2021)]},\
  \Eprint {https://arxiv.org/abs/2008.07108} {arXiv:2008.07108 [astro-ph.CO]}
  \BibitemShut {NoStop}%
\bibitem [{\citenamefont {Clarkson}\ \emph {et~al.}(2011)\citenamefont
  {Clarkson}, \citenamefont {Ellis}, \citenamefont {Larena},\ and\
  \citenamefont {Umeh}}]{bcreview1}%
  \BibitemOpen
  \bibfield  {author} {\bibinfo {author} {\bibfnamefont {C.}~\bibnamefont
  {Clarkson}}, \bibinfo {author} {\bibfnamefont {G.}~\bibnamefont {Ellis}},
  \bibinfo {author} {\bibfnamefont {J.}~\bibnamefont {Larena}},\ and\ \bibinfo
  {author} {\bibfnamefont {O.}~\bibnamefont {Umeh}},\ }\bibfield  {title}
  {\bibinfo {title} {Does the growth of structure affect our dynamical models
  of the universe? the averaging, backreaction, and fitting problems in
  cosmology},\ }\href {https://doi.org/10.1088/0034-4885/74/11/112901}
  {\bibfield  {journal} {\bibinfo  {journal} {Reports on Progress in Physics}\
  }\textbf {\bibinfo {volume} {74}},\ \bibinfo {pages} {112901} (\bibinfo
  {year} {2011})}\BibitemShut {NoStop}%
\bibitem [{\citenamefont {Buchert}(2007)}]{bcreview2}%
  \BibitemOpen
  \bibfield  {author} {\bibinfo {author} {\bibfnamefont {T.}~\bibnamefont
  {Buchert}},\ }\bibfield  {title} {\bibinfo {title} {Dark energy from
  structure: a status report},\ }\href
  {https://doi.org/10.1007/s10714-007-0554-8} {\bibfield  {journal} {\bibinfo
  {journal} {General Relativity and Gravitation}\ }\textbf {\bibinfo {volume}
  {40}},\ \bibinfo {pages} {467–527} (\bibinfo {year} {2007})}\BibitemShut
  {NoStop}%
\bibitem [{\citenamefont {Räsänen}(2011)}]{bcreview3}%
  \BibitemOpen
  \bibfield  {author} {\bibinfo {author} {\bibfnamefont {S.}~\bibnamefont
  {Räsänen}},\ }\bibfield  {title} {\bibinfo {title} {Backreaction:
  directions of progress},\ }\href
  {https://doi.org/10.1088/0264-9381/28/16/164008} {\bibfield  {journal}
  {\bibinfo  {journal} {Classical and Quantum Gravity}\ }\textbf {\bibinfo
  {volume} {28}},\ \bibinfo {pages} {164008} (\bibinfo {year}
  {2011})}\BibitemShut {NoStop}%
\bibitem [{\citenamefont {Mishra}\ \emph
  {et~al.}(2012{\natexlab{b}})\citenamefont {Mishra}, \citenamefont
  {Celerier},\ and\ \citenamefont {Singh}}]{Mishra:2012vi}%
  \BibitemOpen
  \bibfield  {author} {\bibinfo {author} {\bibfnamefont {P.}~\bibnamefont
  {Mishra}}, \bibinfo {author} {\bibfnamefont {M.-N.}\ \bibnamefont
  {Celerier}},\ and\ \bibinfo {author} {\bibfnamefont {T.~P.}\ \bibnamefont
  {Singh}},\ }\bibfield  {title} {\bibinfo {title} {{Redshift drift as a test
  for discriminating between different cosmological models}},\ }\href
  {https://doi.org/10.1103/PhysRevD.86.083520} {\bibfield  {journal} {\bibinfo
  {journal} {Phys. Rev. D}\ }\textbf {\bibinfo {volume} {86}},\ \bibinfo
  {pages} {083520} (\bibinfo {year} {2012}{\natexlab{b}})},\ \Eprint
  {https://arxiv.org/abs/1206.6026} {arXiv:1206.6026 [astro-ph.CO]}
  \BibitemShut {NoStop}%
\bibitem [{\citenamefont {Mishra}\ and\ \citenamefont
  {Celerier}(2022)}]{Mishra:2014vga}%
  \BibitemOpen
  \bibfield  {author} {\bibinfo {author} {\bibfnamefont {P.}~\bibnamefont
  {Mishra}}\ and\ \bibinfo {author} {\bibfnamefont {M.-N.}\ \bibnamefont
  {Celerier}},\ }\bibfield  {title} {\bibinfo {title} {{Redshift and redshift
  drift in $\Lambda=0$ quasispherical Szekeres cosmological models and the
  effect of averaging}},\ }\href {https://doi.org/10.1103/PhysRevD.105.063520}
  {\bibfield  {journal} {\bibinfo  {journal} {Phys. Rev. D}\ }\textbf {\bibinfo
  {volume} {105}},\ \bibinfo {pages} {063520} (\bibinfo {year} {2022})},\
  \Eprint {https://arxiv.org/abs/1403.5229} {arXiv:1403.5229 [astro-ph.CO]}
  \BibitemShut {NoStop}%
\bibitem [{\citenamefont {{Loffler}}\ \emph {et~al.}(2012)\citenamefont
  {{Loffler}}, \citenamefont {{Faber}}, \citenamefont {{Bentivegna}},
  \citenamefont {{Bode}}, \citenamefont {{Diener}}, \citenamefont {{Haas}},
  \citenamefont {{Hinder}}, \citenamefont {{Mundim}}, \citenamefont {{Ott}},
  \citenamefont {{Schnetter}}, \citenamefont {{Allen}}, \citenamefont
  {{Campanelli}},\ and\ \citenamefont {{Laguna}}}]{Loffler:2012}%
  \BibitemOpen
  \bibfield  {author} {\bibinfo {author} {\bibfnamefont {F.}~\bibnamefont
  {{Loffler}}}, \bibinfo {author} {\bibfnamefont {J.}~\bibnamefont {{Faber}}},
  \bibinfo {author} {\bibfnamefont {E.}~\bibnamefont {{Bentivegna}}}, \bibinfo
  {author} {\bibfnamefont {T.}~\bibnamefont {{Bode}}}, \bibinfo {author}
  {\bibfnamefont {P.}~\bibnamefont {{Diener}}}, \bibinfo {author}
  {\bibfnamefont {R.}~\bibnamefont {{Haas}}}, \bibinfo {author} {\bibfnamefont
  {I.}~\bibnamefont {{Hinder}}}, \bibinfo {author} {\bibfnamefont {B.~C.}\
  \bibnamefont {{Mundim}}}, \bibinfo {author} {\bibfnamefont {C.~D.}\
  \bibnamefont {{Ott}}}, \bibinfo {author} {\bibfnamefont {E.}~\bibnamefont
  {{Schnetter}}}, \bibinfo {author} {\bibfnamefont {G.}~\bibnamefont
  {{Allen}}}, \bibinfo {author} {\bibfnamefont {M.}~\bibnamefont
  {{Campanelli}}},\ and\ \bibinfo {author} {\bibfnamefont {P.}~\bibnamefont
  {{Laguna}}},\ }\bibfield  {title} {\bibinfo {title} {{The Einstein Toolkit: a
  community computational infrastructure for relativistic astrophysics}},\
  }\href {https://doi.org/10.1088/0264-9381/29/11/115001} {\bibfield  {journal}
  {\bibinfo  {journal} {Classical and Quantum Gravity}\ }\textbf {\bibinfo
  {volume} {29}},\ \bibinfo {eid} {115001} (\bibinfo {year} {2012})},\ \Eprint
  {https://arxiv.org/abs/1111.3344} {arXiv:1111.3344 [gr-qc]} \BibitemShut
  {NoStop}%
\bibitem [{\citenamefont {{Zilhao}}\ and\ \citenamefont
  {{Loffler}}(2013)}]{Zilhao:2013}%
  \BibitemOpen
  \bibfield  {author} {\bibinfo {author} {\bibfnamefont {M.}~\bibnamefont
  {{Zilhao}}}\ and\ \bibinfo {author} {\bibfnamefont {F.}~\bibnamefont
  {{Loffler}}},\ }\bibfield  {title} {\bibinfo {title} {{An Introduction to the
  Einstein Toolkit}},\ }\href {https://doi.org/10.1142/S0217751X13400149}
  {\bibfield  {journal} {\bibinfo  {journal} {International Journal of Modern
  Physics A}\ }\textbf {\bibinfo {volume} {28}},\ \bibinfo {eid} {1340014-126}
  (\bibinfo {year} {2013})},\ \Eprint {https://arxiv.org/abs/1305.5299}
  {arXiv:1305.5299 [gr-qc]} \BibitemShut {NoStop}%
\bibitem [{\citenamefont {{Bentivegna}}\ and\ \citenamefont
  {{Bruni}}(2016)}]{Bentivegna:2016}%
  \BibitemOpen
  \bibfield  {author} {\bibinfo {author} {\bibfnamefont {E.}~\bibnamefont
  {{Bentivegna}}}\ and\ \bibinfo {author} {\bibfnamefont {M.}~\bibnamefont
  {{Bruni}}},\ }\bibfield  {title} {\bibinfo {title} {{Effects of Nonlinear
  Inhomogeneity on the Cosmic Expansion with Numerical Relativity}},\ }\href
  {https://doi.org/10.1103/PhysRevLett.116.251302} {\bibfield  {journal}
  {\bibinfo  {journal} {\prl}\ }\textbf {\bibinfo {volume} {116}},\ \bibinfo
  {eid} {251302} (\bibinfo {year} {2016})},\ \Eprint
  {https://arxiv.org/abs/1511.05124} {arXiv:1511.05124 [gr-qc]} \BibitemShut
  {NoStop}%
\bibitem [{\citenamefont {{Bentivegna}}(2017)}]{Bentivegna:2017a}%
  \BibitemOpen
  \bibfield  {author} {\bibinfo {author} {\bibfnamefont {E.}~\bibnamefont
  {{Bentivegna}}},\ }\bibfield  {title} {\bibinfo {title} {{Automatically
  generated code for relativistic inhomogeneous cosmologies}},\ }\href
  {https://doi.org/10.1103/PhysRevD.95.044046} {\bibfield  {journal} {\bibinfo
  {journal} {\prd}\ }\textbf {\bibinfo {volume} {95}},\ \bibinfo {eid} {044046}
  (\bibinfo {year} {2017})},\ \Eprint {https://arxiv.org/abs/1610.05198}
  {arXiv:1610.05198 [gr-qc]} \BibitemShut {NoStop}%
\bibitem [{\citenamefont {{Macpherson}}\ \emph {et~al.}(2017)\citenamefont
  {{Macpherson}}, \citenamefont {{Lasky}},\ and\ \citenamefont
  {{Price}}}]{Macpherson:2017}%
  \BibitemOpen
  \bibfield  {author} {\bibinfo {author} {\bibfnamefont {H.~J.}\ \bibnamefont
  {{Macpherson}}}, \bibinfo {author} {\bibfnamefont {P.~D.}\ \bibnamefont
  {{Lasky}}},\ and\ \bibinfo {author} {\bibfnamefont {D.~J.}\ \bibnamefont
  {{Price}}},\ }\bibfield  {title} {\bibinfo {title} {{Inhomogeneous cosmology
  with numerical relativity}},\ }\href
  {https://doi.org/10.1103/PhysRevD.95.064028} {\bibfield  {journal} {\bibinfo
  {journal} {\prd}\ }\textbf {\bibinfo {volume} {95}},\ \bibinfo {eid} {064028}
  (\bibinfo {year} {2017})},\ \Eprint {https://arxiv.org/abs/1611.05447}
  {arXiv:1611.05447 [astro-ph.CO]} \BibitemShut {NoStop}%
\bibitem [{\citenamefont {{Macpherson}}\ \emph {et~al.}(2019)\citenamefont
  {{Macpherson}}, \citenamefont {{Price}},\ and\ \citenamefont
  {{Lasky}}}]{Macpherson:2019a}%
  \BibitemOpen
  \bibfield  {author} {\bibinfo {author} {\bibfnamefont {H.~J.}\ \bibnamefont
  {{Macpherson}}}, \bibinfo {author} {\bibfnamefont {D.~J.}\ \bibnamefont
  {{Price}}},\ and\ \bibinfo {author} {\bibfnamefont {P.~D.}\ \bibnamefont
  {{Lasky}}},\ }\bibfield  {title} {\bibinfo {title} {{Einstein's Universe:
  Cosmological structure formation in numerical relativity}},\ }\href
  {https://doi.org/10.1103/PhysRevD.99.063522} {\bibfield  {journal} {\bibinfo
  {journal} {\prd}\ }\textbf {\bibinfo {volume} {99}},\ \bibinfo {eid} {063522}
  (\bibinfo {year} {2019})},\ \Eprint {https://arxiv.org/abs/1807.01711}
  {arXiv:1807.01711 [astro-ph.CO]} \BibitemShut {NoStop}%
\bibitem [{\citenamefont {{Lesgourgues}}(2011)}]{Lesgourgues:2011}%
  \BibitemOpen
  \bibfield  {author} {\bibinfo {author} {\bibfnamefont {J.}~\bibnamefont
  {{Lesgourgues}}},\ }\bibfield  {title} {\bibinfo {title} {{The Cosmic Linear
  Anisotropy Solving System (CLASS) I: Overview}},\ }\href
  {https://doi.org/10.48550/arXiv.1104.2932} {\bibfield  {journal} {\bibinfo
  {journal} {arXiv e-prints}\ ,\ \bibinfo {eid} {arXiv:1104.2932}} (\bibinfo
  {year} {2011})},\ \Eprint {https://arxiv.org/abs/1104.2932} {arXiv:1104.2932
  [astro-ph.IM]} \BibitemShut {NoStop}%
\bibitem [{\citenamefont {{Macpherson}}(2023)}]{Macpherson:2023}%
  \BibitemOpen
  \bibfield  {author} {\bibinfo {author} {\bibfnamefont {H.~J.}\ \bibnamefont
  {{Macpherson}}},\ }\bibfield  {title} {\bibinfo {title} {{Cosmological
  distances with general-relativistic ray tracing: framework and comparison to
  cosmographic predictions}},\ }\href
  {https://doi.org/10.1088/1475-7516/2023/03/019} {\bibfield  {journal}
  {\bibinfo  {journal} {jcap}\ }\textbf {\bibinfo {volume} {2023}},\ \bibinfo
  {eid} {019} (\bibinfo {year} {2023})},\ \Eprint
  {https://arxiv.org/abs/2209.06775} {arXiv:2209.06775 [astro-ph.CO]}
  \BibitemShut {NoStop}%
\bibitem [{\citenamefont {{Gorski}}\ \emph {et~al.}(2005)\citenamefont
  {{Gorski}}, \citenamefont {{Hivon}}, \citenamefont {{Banday}}, \citenamefont
  {{Wandelt}}, \citenamefont {{Hansen}}, \citenamefont {{Reinecke}},\ and\
  \citenamefont {{Bartelmann}}}]{Gorski:2005}%
  \BibitemOpen
  \bibfield  {author} {\bibinfo {author} {\bibfnamefont {K.~M.}\ \bibnamefont
  {{Gorski}}}, \bibinfo {author} {\bibfnamefont {E.}~\bibnamefont {{Hivon}}},
  \bibinfo {author} {\bibfnamefont {A.~J.}\ \bibnamefont {{Banday}}}, \bibinfo
  {author} {\bibfnamefont {B.~D.}\ \bibnamefont {{Wandelt}}}, \bibinfo {author}
  {\bibfnamefont {F.~K.}\ \bibnamefont {{Hansen}}}, \bibinfo {author}
  {\bibfnamefont {M.}~\bibnamefont {{Reinecke}}},\ and\ \bibinfo {author}
  {\bibfnamefont {M.}~\bibnamefont {{Bartelmann}}},\ }\bibfield  {title}
  {\bibinfo {title} {{HEALPix: A Framework for High-Resolution Discretization
  and Fast Analysis of Data Distributed on the Sphere}},\ }\href
  {https://doi.org/10.1086/427976} {\bibfield  {journal} {\bibinfo  {journal}
  {apj}\ }\textbf {\bibinfo {volume} {622}},\ \bibinfo {pages} {759} (\bibinfo
  {year} {2005})},\ \Eprint {https://arxiv.org/abs/astro-ph/0409513}
  {arXiv:astro-ph/0409513 [astro-ph]} \BibitemShut {NoStop}%
\bibitem [{\citenamefont {{Lema\^{\i}tre}}(1933)}]{lemaitre}%
  \BibitemOpen
  \bibfield  {author} {\bibinfo {author} {\bibfnamefont {G.}~\bibnamefont
  {{Lema\^{\i}tre}}},\ }\bibfield  {title} {\bibinfo {title} {{L'Univers en
  expansion}},\ }\href@noop {} {\bibfield  {journal} {\bibinfo  {journal} {Ann.
  Soc. Sci. Bruxelles A}\ }\textbf {\bibinfo {volume} {53}},\ \bibinfo {pages}
  {51} (\bibinfo {year} {1933})}\BibitemShut {NoStop}%
\bibitem [{\citenamefont {{Tolman}}(1934)}]{tolman}%
  \BibitemOpen
  \bibfield  {author} {\bibinfo {author} {\bibfnamefont {R.~C.}\ \bibnamefont
  {{Tolman}}},\ }\bibfield  {title} {\bibinfo {title} {{Effect of Inhomogeneity
  on Cosmological Models}},\ }\href@noop {} {\bibfield  {journal} {\bibinfo
  {journal} {Proc. Nat. Acad. Sci.}\ }\textbf {\bibinfo {volume} {20}},\
  \bibinfo {pages} {169} (\bibinfo {year} {1934})}\BibitemShut {NoStop}%
\bibitem [{\citenamefont {{Bondi}}(1947)}]{bondi}%
  \BibitemOpen
  \bibfield  {author} {\bibinfo {author} {\bibfnamefont {H.}~\bibnamefont
  {{Bondi}}},\ }\bibfield  {title} {\bibinfo {title} {{Spherically Symmetrical
  Models in General Relativity}},\ }\href@noop {} {\bibfield  {journal}
  {\bibinfo  {journal} {Mon. Not. R. Astr. Soc.}\ }\textbf {\bibinfo {volume}
  {107}},\ \bibinfo {pages} {410} (\bibinfo {year} {1947})}\BibitemShut
  {NoStop}%
\bibitem [{\citenamefont {Rasanen}(2014)}]{Rasanen:2013swa}%
  \BibitemOpen
  \bibfield  {author} {\bibinfo {author} {\bibfnamefont {S.}~\bibnamefont
  {Rasanen}},\ }\bibfield  {title} {\bibinfo {title} {{A covariant treatment of
  cosmic parallax}},\ }\href {https://doi.org/10.1088/1475-7516/2014/03/035}
  {\bibfield  {journal} {\bibinfo  {journal} {JCAP}\ }\textbf {\bibinfo
  {volume} {03}},\ \bibinfo {pages} {035}},\ \Eprint
  {https://arxiv.org/abs/1312.5738} {arXiv:1312.5738 [astro-ph.CO]}
  \BibitemShut {NoStop}%
\bibitem [{\citenamefont {Liske}\ \emph {et~al.}(2008)\citenamefont {Liske}
  \emph {et~al.}}]{Liske:2008ph}%
  \BibitemOpen
  \bibfield  {author} {\bibinfo {author} {\bibfnamefont {J.}~\bibnamefont
  {Liske}} \emph {et~al.},\ }\bibfield  {title} {\bibinfo {title} {{Cosmic
  dynamics in the era of Extremely Large Telescopes}},\ }\href
  {https://doi.org/10.1111/j.1365-2966.2008.13090.x} {\bibfield  {journal}
  {\bibinfo  {journal} {Mon. Not. Roy. Astron. Soc.}\ }\textbf {\bibinfo
  {volume} {386}},\ \bibinfo {pages} {1192} (\bibinfo {year} {2008})},\ \Eprint
  {https://arxiv.org/abs/0802.1532} {arXiv:0802.1532 [astro-ph]} \BibitemShut
  {NoStop}%
\bibitem [{\citenamefont {{Dolag}}\ \emph {et~al.}(2023)\citenamefont
  {{Dolag}}, \citenamefont {{Sorce}}, \citenamefont {{Pilipenko}},
  \citenamefont {{Hern{\'a}ndez-Mart{\'\i}nez}}, \citenamefont {{Valentini}},
  \citenamefont {{Gottl{\"o}ber}}, \citenamefont {{Aghanim}},\ and\
  \citenamefont {{Khabibullin}}}]{Dolag:2023wf}%
  \BibitemOpen
  \bibfield  {author} {\bibinfo {author} {\bibfnamefont {K.}~\bibnamefont
  {{Dolag}}}, \bibinfo {author} {\bibfnamefont {J.~G.}\ \bibnamefont
  {{Sorce}}}, \bibinfo {author} {\bibfnamefont {S.}~\bibnamefont
  {{Pilipenko}}}, \bibinfo {author} {\bibfnamefont {E.}~\bibnamefont
  {{Hern{\'a}ndez-Mart{\'\i}nez}}}, \bibinfo {author} {\bibfnamefont
  {M.}~\bibnamefont {{Valentini}}}, \bibinfo {author} {\bibfnamefont
  {S.}~\bibnamefont {{Gottl{\"o}ber}}}, \bibinfo {author} {\bibfnamefont
  {N.}~\bibnamefont {{Aghanim}}},\ and\ \bibinfo {author} {\bibfnamefont
  {I.}~\bibnamefont {{Khabibullin}}},\ }\bibfield  {title} {\bibinfo {title}
  {{Simulating the LOcal Web (SLOW): I. Anomalies in the local density
  field}},\ }\href {https://doi.org/10.48550/arXiv.2302.10960} {\bibfield
  {journal} {\bibinfo  {journal} {arXiv e-prints}\ ,\ \bibinfo {eid}
  {arXiv:2302.10960}} (\bibinfo {year} {2023})},\ \Eprint
  {https://arxiv.org/abs/2302.10960} {arXiv:2302.10960 [astro-ph.CO]}
  \BibitemShut {NoStop}%
\bibitem [{\citenamefont {{Klypin}}\ \emph {et~al.}(2003)\citenamefont
  {{Klypin}}, \citenamefont {{Hoffman}}, \citenamefont {{Kravtsov}},\ and\
  \citenamefont {{Gottlober}}}]{Klypin:2003uh}%
  \BibitemOpen
  \bibfield  {author} {\bibinfo {author} {\bibfnamefont {A.}~\bibnamefont
  {{Klypin}}}, \bibinfo {author} {\bibfnamefont {Y.}~\bibnamefont {{Hoffman}}},
  \bibinfo {author} {\bibfnamefont {A.~V.}\ \bibnamefont {{Kravtsov}}},\ and\
  \bibinfo {author} {\bibfnamefont {S.}~\bibnamefont {{Gottlober}}},\
  }\bibfield  {title} {\bibinfo {title} {{Constrained Simulations of the Real
  Universe: The Local Supercluster}},\ }\href {https://doi.org/10.1086/377574}
  {\bibfield  {journal} {\bibinfo  {journal} {apj}\ }\textbf {\bibinfo {volume}
  {596}},\ \bibinfo {pages} {19} (\bibinfo {year} {2003})},\ \Eprint
  {https://arxiv.org/abs/astro-ph/0107104} {arXiv:astro-ph/0107104 [astro-ph]}
  \BibitemShut {NoStop}%
\bibitem [{\citenamefont {Heinesen}(2021{\natexlab{d}})}]{Heinesen:2020bej}%
  \BibitemOpen
  \bibfield  {author} {\bibinfo {author} {\bibfnamefont {A.}~\bibnamefont
  {Heinesen}},\ }\bibfield  {title} {\bibinfo {title} {{Multipole decomposition
  of the general luminosity distance 'Hubble law' -- a new framework for
  observational cosmology}},\ }\href
  {https://doi.org/10.1088/1475-7516/2021/05/008} {\bibfield  {journal}
  {\bibinfo  {journal} {Journal of Cosmology and Astroparticle Physics}\
  }\textbf {\bibinfo {volume} {2021}}\bibfield  {number} {\bibinfo  {number} {
  (05)},\ \bibinfo {pages} {008}},\ }\Eprint {https://arxiv.org/abs/2010.06534}
  {arXiv:2010.06534 [astro-ph.CO]} \BibitemShut {NoStop}%
\bibitem [{\citenamefont {{Macpherson}}\ and\ \citenamefont
  {{Heinesen}}(2021)}]{Macpherson:2021a}%
  \BibitemOpen
  \bibfield  {author} {\bibinfo {author} {\bibfnamefont {H.~J.}\ \bibnamefont
  {{Macpherson}}}\ and\ \bibinfo {author} {\bibfnamefont {A.}~\bibnamefont
  {{Heinesen}}},\ }\bibfield  {title} {\bibinfo {title} {{Luminosity distance
  and anisotropic sky-sampling at low redshifts: A numerical relativity
  study}},\ }\href {https://doi.org/10.1103/PhysRevD.104.023525} {\bibfield
  {journal} {\bibinfo  {journal} {\prd}\ }\textbf {\bibinfo {volume} {104}},\
  \bibinfo {eid} {023525} (\bibinfo {year} {2021})},\ \Eprint
  {https://arxiv.org/abs/2103.11918} {arXiv:2103.11918 [astro-ph.CO]}
  \BibitemShut {NoStop}%
\bibitem [{\citenamefont {Grasso}\ and\ \citenamefont
  {Villa}(2022)}]{Grasso:2021iwq}%
  \BibitemOpen
  \bibfield  {author} {\bibinfo {author} {\bibfnamefont {M.}~\bibnamefont
  {Grasso}}\ and\ \bibinfo {author} {\bibfnamefont {E.}~\bibnamefont {Villa}},\
  }\bibfield  {title} {\bibinfo {title} {{BiGONLight: light propagation with
  bilocal operators in numerical relativity}},\ }\href
  {https://doi.org/10.1088/1361-6382/ac35aa} {\bibfield  {journal} {\bibinfo
  {journal} {Class. Quant. Grav.}\ }\textbf {\bibinfo {volume} {39}},\ \bibinfo
  {pages} {015011} (\bibinfo {year} {2022})},\ \Eprint
  {https://arxiv.org/abs/2107.06306} {arXiv:2107.06306 [gr-qc]} \BibitemShut
  {NoStop}%
\bibitem [{\citenamefont {Bolejko}\ \emph {et~al.}(2005)\citenamefont
  {Bolejko}, \citenamefont {Krasiński},\ and\ \citenamefont
  {Hellaby}}]{Bolejko_2005}%
  \BibitemOpen
  \bibfield  {author} {\bibinfo {author} {\bibfnamefont {K.}~\bibnamefont
  {Bolejko}}, \bibinfo {author} {\bibfnamefont {A.}~\bibnamefont
  {Krasiński}},\ and\ \bibinfo {author} {\bibfnamefont {C.}~\bibnamefont
  {Hellaby}},\ }\bibfield  {title} {\bibinfo {title} {Formation of voids in the
  universe within the lemaitre-tolman model},\ }\href
  {https://doi.org/10.1111/j.1365-2966.2005.09292.x} {\bibfield  {journal}
  {\bibinfo  {journal} {Monthly Notices of the Royal Astronomical Society}\
  }\textbf {\bibinfo {volume} {362}},\ \bibinfo {pages} {213–228} (\bibinfo
  {year} {2005})}\BibitemShut {NoStop}%
\bibitem [{\citenamefont {{Adamek}}\ \emph {et~al.}(2020)\citenamefont
  {{Adamek}}, \citenamefont {{Barrera-Hinojosa}}, \citenamefont {{Bruni}},
  \citenamefont {{Li}}, \citenamefont {{Macpherson}},\ and\ \citenamefont
  {{Mertens}}}]{Adamek:2020}%
  \BibitemOpen
  \bibfield  {author} {\bibinfo {author} {\bibfnamefont {J.}~\bibnamefont
  {{Adamek}}}, \bibinfo {author} {\bibfnamefont {C.}~\bibnamefont
  {{Barrera-Hinojosa}}}, \bibinfo {author} {\bibfnamefont {M.}~\bibnamefont
  {{Bruni}}}, \bibinfo {author} {\bibfnamefont {B.}~\bibnamefont {{Li}}},
  \bibinfo {author} {\bibfnamefont {H.~J.}\ \bibnamefont {{Macpherson}}},\ and\
  \bibinfo {author} {\bibfnamefont {J.~B.}\ \bibnamefont {{Mertens}}},\
  }\bibfield  {title} {\bibinfo {title} {{Numerical solutions to Einstein's
  equations in a shearing-dust universe: a code comparison}},\ }\href
  {https://doi.org/10.1088/1361-6382/ab939b} {\bibfield  {journal} {\bibinfo
  {journal} {Classical and Quantum Gravity}\ }\textbf {\bibinfo {volume}
  {37}},\ \bibinfo {eid} {154001} (\bibinfo {year} {2020})},\ \Eprint
  {https://arxiv.org/abs/2003.08014} {arXiv:2003.08014 [astro-ph.CO]}
  \BibitemShut {NoStop}%
\end{thebibliography}%

\end{document}